\documentclass[
 ,draft            
  ]
 {aipproc}
\pdfoutput=1
\usepackage{epstopdf}
\usepackage{epsfig}
\usepackage{amsmath}
\usepackage{units}
\usepackage{graphicx}  
\usepackage{braket}
\usepackage{xfrac}
\usepackage[normalem]{ulem}
\usepackage{booktabs}
\usepackage{tabularx}
\newcommand{\op}[1]{\ensuremath{\hat{#1}}}

\newcolumntype{Y}{>{\centering\arraybackslash}X}

\layoutstyle{8x11single}

\begin{document}
\setcounter{page}{1}
\title{Review of Student Difficulties in Upper-Level Quantum Mechanics}


\author{Chandralekha Singh}{
  address={Department of Physics and Astronomy, University of Pittsburgh, Pittsburgh, PA 15260}
}

\author{Emily Marshman}{
}

\begin{abstract}

Learning advanced physics, in general, is challenging not only due to the increased mathematical sophistication but also because one must continue to build on all of the prior knowledge acquired at the introductory and intermediate levels. In addition, learning quantum mechanics can be especially challenging because the paradigms of classical mechanics and quantum mechanics are very different. Here, we review research on student reasoning difficulties in learning upper-level quantum mechanics and research on students' problem-solving and metacognitive skills in these courses. Some of these studies were multi-university investigations. The investigations suggest that there is large diversity in student performance in upper-level quantum mechanics regardless of the university, textbook, or instructor, and many students in these courses have not acquired a functional understanding of the fundamental concepts. The nature of reasoning difficulties in learning quantum mechanics is analogous to reasoning difficulties found via research in introductory physics courses. The reasoning difficulties were often due to over-generalizations of concepts learned in one context to another context where they are not directly applicable. Reasoning difficulties in distinguishing between closely related concepts and in making sense of the formalism of quantum mechanics were common. We conclude with a brief summary of the research-based approaches that take advantage of research on student difficulties in order to improve teaching and learning of quantum mechanics.

\end{abstract}

\maketitle

\section{Introduction}

\subsection{Learning in Upper-level Physics vs. Introductory Physics}

Helping students learn to ``think like a physicist'' is a major goal of many physics courses from the introductory to the advanced level~\cite{arons,lillian,mcdermott1,mcdermott2,reif,reif2,singh2001,loverude,chasteen}. 
In order to become an expert in physics, the development of problem-solving, reasoning, and metacognitive skills must go hand-in-hand with learning content and building a robust knowledge structure~\cite{mcdermott2,reif,reif2,simon,anderson,ericsson}. Expert physicists monitor their own learning and use problem solving as an opportunity for learning, repairing, extending, and organizing their knowledge structure. Much research in physics education has focused on investigating students' reasoning difficulties in learning introductory physics and on the development of research-based curricula and pedagogies that can significantly reduce these difficulties and help students develop a robust knowledge structure~\cite{mcdermott1,mcdermott2}. A parallel strand of research in introductory physics has focused on how a typical student in such courses differs from a physics expert and the strategies that may help students become better problem solvers and independent learners~\cite{reif,reif2}. However, relatively few investigations have focused on the nature of expertise of upper-level physics students and strategies that can be effective in such courses to help students learn physics and develop their problem-solving, reasoning, and higher-order thinking skills further~\cite{mason1,shihyin}.

Learning physics is challenging even at the introductory level because it requires drawing meaningful inferences and unpacking and applying the few fundamental physics principles, which are in compact mathematical forms, to diverse situations~\cite{mcdermott1,mcdermott2}. Learning upper-level physics is also challenging because one must continue to build on all of the prior knowledge acquired at the introductory and intermediate levels. In addition, the mathematical sophistication required is generally significantly higher for upper-level physics. In order to develop a functional understanding, students must focus on the physics concepts while solving problems and be able to go back and forth between the mathematics and the physics, regardless of whether they are converting a physical situation to a mathematical representation or contemplating the physical significance of the result of a complex mathematical procedure during problem solving. However, little is actually known about how expertise in physics develops as a student makes a transition from introductory to intermediate to advanced physics courses and whether the cognitive and metacognitive skills~\cite{metacognition} of advanced students are significantly superior to those of physics majors in the introductory and intermediate level courses. In particular, there is a lack of research on whether the development of these skills from the introductory level to the point at which the students take up scientific careers is a continuous process of development or whether there are some discontinuous boosts in this process, for example, when students become involved in undergraduate or graduate research or when they independently start teaching and/or researching. There are also little research data on what fraction of students who have gone through the entire traditional physics curriculum including the upper-level courses have developed sufficient cognitive and metacognitive skills to excel professionally in the future, e.g., in graduate school or a future career. Investigations in which students in advanced physics courses are asked to perform tasks related to simple introductory physics content cannot properly assess their learning and self-monitoring skills~\cite{metacognition}. Advanced students may possess a large amount of compiled knowledge about introductory physics due to repetition of the basic content in various courses and may not need to do much self-monitoring while solving introductory problems. Therefore, the task of evaluating upper-level students' learning and self-monitoring skills should involve physics topics at the periphery of their own understanding. 
   
\subsection{Effect of the ``Paradigm Shift'' on Student Difficulties in Quantum Mechanics}

Among upper-level courses, quantum mechanics can be especially challenging for students because the paradigms of classical mechanics  and quantum mechanics are very different~\cite{kuhn,griffiths}. For example, unlike classical physics, in which position and momentum are deterministic variables, in quantum mechanics they are operators that act on a wavefunction (or a state) which lies in an abstract Hilbert space. In addition, according to the Copenhagen interpretation, which is most commonly taught in quantum mechanics courses, an electron in a hydrogen atom does not, in general, have a definite distance from the nucleus; it is the act of measurement that collapses the wavefunction and makes it localized at a certain distance. If the wavefunction is known right before the measurement, quantum theory only provides the probability of measuring the distance in a narrow range.
   
The significantly different paradigms of classical mechanics and quantum mechanics suggest that even students with a good knowledge of classical mechanics will start as novices and gradually build their knowledge structure about quantum mechanics. The ``percolation model'' of expertise can be particularly helpful in knowledge-rich domains such as physics~\cite{simon}. In this model of expertise, a person's long term memory contains different ``nodes'' which represent different knowledge pieces within a particular knowledge domain. Experts generally have their knowledge hierarchically organized in pyramid-shaped schema in which the top nodes are more foundational than nodes at a lower level and nodes are connected to other nodes through links that signify the relation between those concepts. As a student develops expertise in a domain, links are formed which connect different knowledge nodes. If a student continues her effort to organize, repair, and extend her knowledge structure, she will reach a percolation threshold when all knowledge nodes become connected to each other by at least one link in an appropriate manner. At this point, the student will become at least a nominal expert. The student can continue on her path to expertise with further strengthening of the nodes and building additional appropriate links. Redundancy in appropriate links between different nodes is useful because it provides alternative pathways during problem solving when other pathways cannot be accessed, e.g., due to memory decay. As a student starts to build a knowledge structure about quantum mechanics, her knowledge nodes will not be appropriately connected to other nodes farther away, and her reasoning about quantum mechanics will only be locally consistent and lack global consistency~\cite{disessa1,disessa2}. In fact, a person who begins a pursuit of expertise in any knowledge-rich domain must go through a phase in which her knowledge is in small disconnected pieces which are only locally consistent but lack global consistency, leading to reasoning difficulties. Therefore, introductory students learning classical mechanics and advanced students learning quantum mechanics are likely to show similar patterns of reasoning difficulties as they move up along the expertise spectrum in each of these sub-domains of physics. 

\subsection{Overview of Student Difficulties in Quantum Mechanics}

Students taking upper-level quantum mechanics often develop survival strategies for performing reasonably well in their course work. For example, they become proficient at solving algorithmic problems such as the time-independent Schr\"odinger equation with a complicated potential energy and boundary conditions. However, research suggests that they often struggle to make sense of the material and build a robust knowledge structure. They have difficulty mastering concepts and applying the formalism to answer qualitative questions, e.g., questions related to the properties of wavefunctions, possible wavefunctions for a given system, the time-development of a wavefunction, measurement of physical observables within the Copenhagen interpretation, and the meaning of expectation values as an ensemble average of a large number of measurements on identically prepared systems~\cite{brad,sadaghiani,guangtian,singh2001,singhtoday,singhgrad}.

Here, we review research on student reasoning difficulties and on their problem-solving and metacognitive skills in learning upper-level quantum mechanics. 
Difficulties in learning quantum mechanics can result from its novel paradigm, the abstractness of the subject matter, and mathematical sophistication. Also, the diversity in students' prior preparation for upper-level courses such as quantum mechanics has increased significantly~\cite{aip} and makes it difficult for instructors to target instruction at the appropriate level. Moreover, in order to transfer previous learning, e.g., knowledge of linear algebra, waves, or probability concepts learned in other contexts, students must first learn the basic structure of quantum mechanics and then contemplate how the previously learned knowledge applies to this novel framework~\cite{gick}. Research suggests that students in upper-level quantum mechanics have common difficulties independent of their background, teaching style, textbook, and institution that are analogous to the patterns of difficulties observed in introductory physics courses, and many students in these courses have not acquired a functional understanding of the fundamental concepts~\cite{singh2001,singhgrad}. The  nature of conceptual difficulties in learning quantum mechanics is analogous in nature to conceptual difficulties found via research in introductory physics courses. 

Several investigations have strived to improve the teaching and learning of quantum  mechanics at the introductory or intermediate level~\cite{zollman,zollman2,redish,redish2,narst,bailey,fischler,ireson,niedderer,muller,lawless,mckagan,photo,bohrmodel,phet,tunnel,qmcs}.
For example, some investigations have focused on students' conceptions about modern physics early in college or at the pre-college level~\cite{fischler,ireson,niedderer,muller,lawless}. Zollman et al.~\cite{zollman} have proposed that quantum concepts be introduced much earlier in physics course sequences and have designed tutorials and visualization tools~\cite{visual} which illustrate concepts that can be used at a variety of levels. 
Redish et al.~\cite{redish,redish2} have conducted investigations of student difficulties and developed research-based material to teach quantum mechanics concepts to a wide range of science and engineering students. Robinett et al.~\cite{robinett} designed a ``visualization'' test related to quantum physics concepts that can be 
administered to students in introductory quantum physics. Other visualization tools have also been developed to help students learn quantum mechanics better~\cite{mario,physlet,brandt,thaller,hiller,antje}.

While there is overlap between the content in introductory, intermediate, and upper-level quantum mechanics courses, here we focus only on student difficulties in upper-level (junior/senior level) quantum mechanics. We first describe theoretical frameworks that inform why investigations of student difficulties in learning quantum mechanics are important. Then, we summarize the methodologies used in the investigations that explore the difficulties. We then  present a summary of common student difficulties in learning upper-level quantum mechanics found via research. We conclude with a brief summary of research-based learning approaches that take into account the research on student difficulties and  strive to help students develop a good knowledge structure of quantum mechanics.

\section{Theoretical frameworks that inform the investigations on student difficulties}

Research on student reasoning difficulties in learning upper-level quantum mechanics and on students' problem-solving and metacognitive skills in these courses is inspired by cognitive theories that point to the importance of knowing student difficulties in order to help them develop a functional understanding of relevant concepts. For example, Hammer proposed a ``resource'' model that suggests that students' prior knowledge, including their learning difficulties, should be used as a resource to help students learn better~\cite{hammer}. Similarly, the Piagetian model of learning emphasizes an ``optimal mismatch'' between what the student knows and is able to do and the instructional design~\cite{piaget,posner}. In particular, this model focuses on the importance of knowing students' skill levels and reasoning difficulties and using this knowledge to design instruction to help them assimilate and accommodate new ideas and build a good knowledge structure. Similarly, Bransford and Schwartz's framework, ``preparation for future learning'' (PFL), suggests that to help students be able to transfer their knowledge from one context to another, instructional design should include elements of both innovation and efficiency~\cite{schwartz}. While there are multiple interpretations of their model, efficiency and innovation can be considered two orthogonal dimensions in instructional design. If instruction only focuses on efficiently transferring information, cognitive engagement will be diminished and learning will not be effective. On the other hand, if the instruction is solely focused on innovation, students will struggle to connect what they are learning with their prior knowledge and learning and transfer will be inhibited. Incorporating the efficiency and innovation elements into an instructional design based upon this framework and being in the ``optimal adaptability corridor'' demands that instruction build on students' existing skills and take into account their reasoning difficulties. With this knowledge for a given student population, an instructor can determine what is innovative and what is efficient.  Vygotsky developed a theory which introduces the notion of the ``zone of proximal development'' (ZPD). The ZPD refers to the zone defined by the difference between what a student can do on his/her own and what a student can do with the help of an instructor who is familiar with his/her prior knowledge and skills~\cite{vygotsky}. Scaffolding is at the heart of this model and can be used to stretch students' learning beyond their current knowledge by carefully crafted instruction. Even within this model of learning, knowing the ZPD requires knowledge of student reasoning difficulties and the current level of expertise in their problem-solving, reasoning, and self-monitoring skills. These cognitive theories (i.e., ``resource'' model, ``optimal mismatch'' model, PFL model, and Vygotsky's model focusing on ZPD) all point to the fact that one must determine the initial knowledge states of the students in order to design effective instruction commensurate with students' current knowledge and skills. Thus, the investigation of student difficulties, which is reviewed in the following sections, can help in the development of curricula and pedagogies to reduce the difficulties and improve learning of quantum mechanics.

\section{Summary of Methodology}

The research studies on learning difficulties in upper-level quantum mechanics summarized in this report use both quantitative and qualitative methodologies. In fact, almost all of the investigations that we draw upon use a mixed methodology involving both quantitative and qualitative data. The complete details of the methodologies can be found in the respective references. However, generally, for the quantitative part of the studies, students in various upper-level undergraduate quantum mechanics courses (after traditional instruction) or in various graduate core quantum mechanics courses (before instruction) were given written surveys with free-response and/or multiple-choice questions on topics that are covered in a typical undergraduate quantum mechanics course.  
Some of these studies were conducted at several universities simultaneously (with the total number of students varying from close to a hundred to more than two hundred depending upon the investigation) while others were conducted at typical state universities where the student population in the upper-level quantum mechanics courses is likely to be representative of students in similar courses at other typical state universities. 

In most studies (which used a mixed research methodology), a subset of students (a smaller number of students than in the quantitative classroom investigations involving written tasks) were interviewed to investigate difficulties with quantum mechanics concepts in more depth and to unravel the underlying cognitive mechanisms. For these qualitative studies, upper-level undergraduate students in various quantum mechanics courses and physics graduate students who were taking or had taken core graduate level quantum mechanics were interviewed individually outside of the class using semi-structured, think-aloud interviews~\cite{chi} and were asked to solve similar problems to those that were administered in written tests. As noted, the rationale was to understand the cognitive mechanism for students' written responses in-depth. In these semi-structured interviews, students  were asked to verbalize their thought processes while they worked on the problems. They were not disturbed while they answered the questions except when asked to ``keep talking'' if they became quiet for a long time. After the students had answered the questions to the best of their ability, they were asked for clarifications of points they had not made clear earlier. In some interviews, students were also asked about their problem solving and learning strategies and what difficulties they faced in learning quantum mechanics. These interviews were semi-structured in the sense that the interviewers had a list of issues that they definitely wanted to discuss. These issues were not brought up initially because the researchers wanted to give students an opportunity to articulate their thought processes and formulate their own responses. However, some of the later probing questions were from the list of issues that researchers had planned to discuss ahead of time (and interviewers asked students at the end of the interview if students did not bring the issue up themselves). Other probing questions were designed on-the-spot by the interviewer to get a better comprehension of a particular student's reasoning and thought process. We note that in some investigations, the individual interview protocol was somewhat different and can be found by consulting the individual references.

\section{Student reasoning difficulties in upper-level quantum mechanics}

Learning content and development of skills go hand in hand.  This section focuses on student reasoning difficulties with different topics in upper-level quantum mechanics and the next section focuses on evidence that students in these courses often have inadequate problem-solving, reasoning, and metacognitive skills.

Similar to research in introductory physics learning, research in learning quantum mechanics suggests that student reasoning difficulties are often context dependent. In other words, a student reasoning difficulty related to a particular topic may manifest itself in one context but not in another context.
This is expected because students are developing expertise and their knowledge structure is not robust. They may recognize the relevance of a particular principle or concept in one context but not in another. Moreover, even students who have a good knowledge structure of mathematics may have conceptual difficulties, especially in a traditional course that focuses mostly on algorithmic problems rather than on sense making.

Furthermore, student responses are sensitive to the wording of a question, particularly for multiple choice questions which include an explicit mention of a particular difficulty. For example, 
students were asked if ``$\hat{H} \Psi =E \Psi$'' is true for all possible wavefunctions for a system (where $\hat{H}$ and $\Psi$ are the Hamiltonian and wave function, respectively). About {\it one out of ten} students incorrectly claimed that it is not true because, instead, the Hamiltonian acting on a generic state corresponds to energy measurement and implies that ``$\hat{H} \Psi =E_n \phi_n$''~\cite{singhtoday}.
On the other hand, when students are explicitly asked to evaluate the correctness of the statement that ``$\hat{H} \Psi =E_n \phi_n$ is true because the Hamiltonian operator acting on a generic state corresponds to the measurement of energy which collapses the state to an energy eigenstate $\phi_n$ and the corresponding energy eigenvalue $E_n$ is measured,'' {\it more than one-third} of students incorrectly agree with this statement~\cite{analogous}. The difference between the percentages of students in these contexts is mainly due to the fact that in one case, students may generate the incorrect expression  ``$\hat{H} \Psi =E_n \phi_n$'' themselves, whereas in the other case they are evaluating the correctness of a statement that explicitly involves 
``$\hat{H} \Psi =E_n \phi_n$.'' This type of context dependence of student responses should be kept in mind in the research studies discussed below. In particular, even if only $5-10\%$ of the students show a certain type of difficulty in a particular context, it is likely that a higher percentage will display the same difficulty in a different context.

We also note that in several studies, very similar problems were chosen to probe student reasoning difficulties in upper-level quantum mechanics. In some cases, the contexts of the problems in two different investigations were very similar except that one study asked students to solve a problem in an open-ended format while the other study asked them to solve the same problem in a multiple-choice format. If there are several contexts in which reasoning difficulties related to a particular topic were investigated, we only present a few examples to illustrate the main issues involved. The original references should be consulted for further details.

\subsection{Difficulties in reconciling quantum concepts with classical concepts}

Quantum mechanics is abstract and its paradigm is very different from the classical paradigm. A good grasp of the principles of quantum mechanics
requires building a knowledge structure consistent with the quantum postulates. However, students often have difficulty reconciling classical concepts with quantum concepts. For example, the fact that measurements are probabilistic and position and momentum do not have the usual meaning in quantum mechanics is very difficult for students. While there are many examples that fall in this broad category of student difficulties in reconciling quantum concepts with classical concepts, here we give a few examples.\\

\textbf{Incorrect belief that a particle loses energy in quantum tunneling}:
Students  have difficulty with the concept of quantum tunneling. Research has shown that students often transfer classical reasoning when thinking about quantum tunneling~\cite{wittman1,wittman2}.  Many students state that a particle ``loses energy'' when it tunnels through a rectangular potential barrier. This reasoning is incorrect because the particle does not lose energy when tunneling through the barrier, although the wave function of the particle inside the potential barrier is described by exponential decay. In interview situations, common responses regarding tunneling involve statements such as:  ``the particle collides and loses energy in the barriers" and ``it requires energy to go through the barrier''~\cite{wittman1,wittman2}. These types of responses indicate that many students incorrectly apply classical concepts to quantum mechanical situations.\\

\textbf{Difficulties distinguishing between a quantum harmonic oscillator vs. a classical harmonic oscillator}:
In one investigation, students had difficulty with the fact that for a simple quantum harmonic oscillator in the ground state, the probability of finding the particle is maximum at the center of the well. For a classical harmonic oscillator, e.g., a simple pendulum, the particle is more likely to be found close to the classical turning points~\cite{zhuqms,zhuqms1}. Discussions with individual students suggest that this difficulty often has its origin in their experiences with how much time a particle spends near the turning points in a classical system.\\

\textbf{Incorrect belief that quantities with  labels ``$x$,'' ``$y$,'' and ``$z$'' are orthogonal to each other}:
One common difficulty upper-level students in quantum mechanics courses have is assuming that an object with a label
``$x$'' is orthogonal to or cannot influence an object with a label ``$y$''~\cite{singhtransfer,sga,my2}. This is evident from responses such as: ``The magnetic field is in the
$z$-direction so the electron is not influenced if it is initially in an eigenstate of $\hat S_x$'' or ``Eigenstates of $\hat S_x$ are orthogonal to eigenstates
of $\hat S_y$.'' In introductory physics, $x$, $y$ and $z$ are indeed conventional labels for orthogonal components of a vector. Unless students are given
an opportunity to understand the structure of quantum mechanics and that the eigenstates of spin components are vectors in Hilbert space and not the physical space in which the magnetic field is a vector, such difficulties will persist. Students must learn that although an electron in an external magnetic field pointing in the $z$-direction is in a real, physical, three-dimensional space of the laboratory, making predictions about the measurement performed in the laboratory using quantum mechanics requires mapping the problem to an abstract Hilbert space in which the state of the system lies and
where all the observables of the real physical space get mapped onto operators acting on states.\\

\textbf{Difficulties with photon polarization states}:
In an investigation involving photon polarization states, some interviewed students claimed that the polarization states of a photon cannot be used as basis vectors for a two-state system due to the fact that a photon can have an infinite number of polarization states~\cite{sethqkd,mz1,mz2}. They argued that since a polarizer can have any orientation and the orientation of the polarizer determines the polarization state of a photon after it passes through the polarizer, it did not make sense to think about the polarization states of a photon as a two-state system. These students were often so fixated on their experiences with polarizers from introductory physics courses (which can be rotated to make their polarization axis along any direction perpendicular to the direction of propagation) that they had difficulty thinking about the polarization states of a photon as vectors in a two-dimensional space. 
It is interesting to note that most students who had difficulty accepting that the polarization states of a photon can be used as basis states for a two-state system had no difficulty accepting that spin states of a spin-1/2 particle can be used as basis states for a two-state system despite the fact that these two systems are isomorphic from an expert perspective. Interviews suggest that this difference in their perception was often due to how a spin-1/2 system and polarization were first introduced and the kinds of mental models students had built about each system. Generally, students are introduced to polarization in an introductory course and to spin-1/2 systems in a quantum mechanics course. Discussions suggest that some students were so used to thinking about a beam of light passing through a polarizer according to their own mental model that they had difficulty thinking about the polarization states of a photon as vectors in a two-dimensional Hilbert space. Many instructors introduce polarization basis vectors in classical electricity and magnetism, but many students do not remember these concepts. Since students had learned about the spin-1/2 system only in quantum mechanics, thinking of the spin states of a spin-1/2 particle as vectors in a two-dimensional space often did not create a similar conflict.\\
   
\textbf{Difficulties with the wave-particle duality}:   
The double-slit experiment reveals that the wavefunction of a single electron can be non-zero through both slits.
In particular, if electrons are sent one at a time through two slits, under appropriate conditions, one observes an interference pattern after a large number of electrons have arrived on a distant phosphor screen. This experiment is very difficult to reconcile with classical ideas.
While the wavefunction of a single electron is non-zero through both slits, when the electron arrives at a detecting screen, a flash is seen in one location due to the collapse of the wave function. The wave-particle duality of a single electron, which is evident at different times in the same experiment, is very difficult for students to rationalize~\cite{griffiths,brad,narst}. Students may have used vocabulary such as ``particle'' to describe a localized entity in their classical mechanics courses. Consequently, they may find it very difficult to think of the electron as a wave in part of the experiment (when it is going through the two slits) and as a particle in another part of the experiment (when it lands on the detecting screen and the wavefunction collapses). 

\subsection{Difficulties with the wavefunction}

Any smooth, normalized function that satisfies the boundary conditions for a system is a possible wavefunction. However, students struggle to determine possible wavefunctions, especially if they are not explicitly written as a linear superposition of stationary states. The following difficulties have been found via research~\cite{singhgrad,zollman2,zhuqms1,guang2}:\\

\textbf{Incorrect belief that ``$\hat H \Psi=E \Psi$'' holds for any possible wavefunction $\Psi$}:
In a multi-university study~\cite{singhgrad}, many students claimed that the Time-Independent Schr\"odinger Equation (TISE)
$\hat H \Psi=E \Psi$ is true for all possible wavefunctions, even when $\Psi$  is not an energy eigenstate (stationary state). In general, $\Psi=\sum_{n=1}^{\infty} C_n \phi_n$, where $\phi_n$ are the stationary
 states and $C_n=\langle \phi_n \vert \Psi \rangle$. Therefore, $\hat H \Psi=\sum_{n=1}^{\infty} C_n E_n \phi_n \ne E \Psi$. More than one-third of the students incorrectly stated that the expression ``$\hat H \Psi=E \Psi$'' is unconditionally
correct, with statements such as the following being typical:
``Agree. This is what 80 years of experiment has proven. If future experiments prove this statement wrong,
then I'll update my opinion on this subject.'' Students with such responses misunderstood what the instructor taught, perhaps due to an overemphasis on the TISE in the course. This incorrect notion that all possible wavefunctions should satisfy the TISE makes it challenging for students to determine possible wavefunctions for a given system. 
In individual interviews, students were explicitly asked whether ``$\hat H \Psi=E \Psi$'' is true for a linear superposition of the ground and first excited state wavefunctions, $\phi_1$ and $\phi_2$, respectively, for a one-dimensional infinite square well. Many students incorrectly claimed that ``$\hat H \Psi=E \Psi$'' is indeed true for this wavefunction. When these students were asked to explicitly show that this equation is true in this given context, most of them verbally argued without writing that  since 
the TISE works for each $\phi_1$ and $\phi_2$ individually, it  implies that it should be satisfied by their linear superposition. In fact, even when students were told that the TISE is not satisfied for this linear superposition, many had difficulty believing it until they explicitly wrote these equations on paper (mostly after additional encouragement to do so) and noted that since $E_1$ and $E_2$ are not equal, $\hat H \Psi \ne E \Psi$ in this case.\\

\textbf{Difficulties with mathematical representations of non-stationary state wavefunctions}:
Students have difficulties in determining non-stationary possible wavefunctions for a given quantum system. For example, in a multi-university study in Ref.~\cite{singhgrad}, student interviewees were given three wavefunctions and asked if they were possible wavefunctions for an electron in a one-dimensional infinite square well between $x=0$ and $x=a$ and to explain their reasoning. Students had to note that the first wavefunction $A e^{-{((x-a/2)/a)}^2}$ is not possible because it does not satisfy the boundary conditions (does not go to zero at $x=0$ and $x=a$). The other two wavefunctions,
$A\sin^3(\pi x/a)$ and $A[\sqrt{2/5} \sin(\pi x/a)+ \sqrt{3/5} \sin(2\pi x/a)]$ with suitable normalization constants, are both smooth functions that satisfy the boundary conditions (each of them goes to zero at $x=0$ and $x=a$). Thus, each can be written as a linear superposition of the stationary states. More than three-fourths of the students could identify that the wavefunction written as a linear combination is a possible wavefunction because it was explicitly written in the form of a linear superposition of stationary states but only
one-third gave the correct answer for all three wavefunctions. About half of the students claimed that $A\sin^3(\pi x/a)$ is not a possible wavefunction but that $A[\sqrt{2/5} \sin(\pi x/a)+ \sqrt{3/5} \sin(2\pi x/a)]$ is possible.
The interviews suggest that a majority of students did not know that any smooth, single-valued wavefunction that satisfies the boundary
conditions can be written as a linear superposition of stationary states.
Interviews and written explanations also suggest that many students incorrectly thought that any possible wavefunction must satisfy both of the following constraints: 1) it must be a smooth, single-valued function that satisfies the boundary conditions; and 2) it must either be possible to write it as a linear superposition of stationary states or it must satisfy the Time-Independent
Schr\"odinger Equation. Some students who correctly realized that $A\sin^3(\pi x/a)$ satisfies the boundary conditions
incorrectly claimed that it is still not a possible wavefunction because it does not satisfy the TISE.
Many students claimed that only pure sinusoidal wavefunctions are possible, thus functions involving $\sin^2$ or $\sin^3$ are not possible wavefunctions.
Many students thought that $A\sin^3(\pi x/a)$ cannot be written as a
linear superposition of stationary states and hence it is not a possible wavefunction while others claimed that
$A\sin^3(\pi x/a)$ works for three electrons but not one.\\

\textbf{Difficulties with diverse representations of a wavefunction}:
In another multi-university investigation, students were given a valid and reliable survey with multiple-choice questions~\cite{zhuqms1}. On one question, graphs (or diagrams) of three possible wavefunctions for a one-dimensional infinite square well were provided in which two graphs displayed stationary state wavefunctions and one showed a non-stationary state wavefunction. All wavefunctions were possible because they were smooth and satisfied the boundary conditions for a one-dimensional infinite square well. Students were asked to choose all wavefunctions that are possible for the infinite square well. In response, half of the students incorrectly claimed that only the stationary state wavefunctions are possible. On the same survey, more than one-third of the students incorrectly claimed that a possible wavefunction must be an even or odd function if the potential energy is a symmetric function  due to a confusion with the energy eigenstates for familiar problems. Also, on another question on the same survey, many students correctly noted that a linear superposition of stationary states is a possible wavefunction for a one-dimensional infinite square well. However, students did not answer different questions about the same system consistently. In particular, many students who noted that the possible wavefunction must be an even or odd function if the potential energy is a symmetric function also noted that a linear superposition of stationary states is a possible wavefunction for a one-dimensional infinite square well which is contradictory since a linear combination of energy eigenstates for this system is not necessarily an even or odd function. Similarly, those who only selected even or odd functions as possible wavefunctions in the diagrammatic representation often claimed that a linear superposition of stationary states is a possible wavefunction for a one-dimensional infinite square well, which is also contradictory.

On the same survey~\cite{zhuqms1}, in the context of a finite square well,  students were given diagrammatic representation of a possible wavefunction which is non-zero only in the well (it goes to zero outside the well), and they were asked if it is a possible wavefunction. Less than half correctly identified it as a possible wavefunction for a finite square well. More than half incorrectly claimed that it is not a possible wavefunction because it does not satisfy the boundary conditions (it goes to zero inside the well) and the probability of finding the particle outside the finite square well is zero but quantum mechanically it must be nonzero. Thus, many students incorrectly thought that any possible wavefunction for a finite square well must have a non-zero probability in the classically forbidden region.\\ 

\textbf{Difficulties with bound states and scattering states}:
When a quantum particle is in an energy eigenstate or a superposition of energy eigenstates such that the energy is less than the potential energy at both plus and minus infinity, the particle is in a bound state. Otherwise, 
it is in a scattering state. Here, we will only discuss situations in which the bound states and scattering states refer to stationary states since most investigations of student difficulties have focused on those cases. The bound states have a discrete energy spectrum and the scattering states have a continuous energy spectrum. Bound state wavefunctions go to zero at infinity so they can always be normalized. Scattering state wavefunctions are not normalizable since the probability of finding the particle is non-zero at infinity, but a normalized wavefunction can be constructed using their linear superpositions. 
 
Students have difficulties with various aspects of the bound and scattering states of a quantum system~\cite{zollman2,zhuqms1,guang2}. In a multi-university survey~\cite{zhuqms1}, on questions focusing on students' knowledge about the bound and scattering state wavefunctions, many students either claimed that the scattering state wavefunctions are normalizable or they did not recognize that a linear superposition of the scattering state wavefunctions can be normalized. Moreover, more than one-third did not recognize that the scattering states have a continuous energy spectrum and claimed that energy is always discrete in quantum mechanics while a comparable percentage of the students claimed that the finite square well only allows discrete energy states (bound states).
  
On several questions on the same survey that required students to judge whether a given potential energy  allows for bound states or scattering states, students had great difficulties~\cite{zhuqms1}. One question uses a graphical representation showing four different potential energy wells. The distractor (incorrect answer) that the students found challenging was a graph in which the potential energy of the well bottom was greater than the potential energy at infinity (which is zero). Therefore, no bound state can exist in this potential energy well. About two-thirds of the students failed to interpret these features. They thought that any potential energy  that has the shape of a ``well'' would allow for bound states if there were classical turning points. 
  
Some questions on the survey focused on the common student difficulty that a given quantum particle may be in a bound or a scattering state depending on its location. This notion often has its origin in students' classical experiences. In particular, some students mistakenly claimed that a particle could have different energies in different regions in a potential energy diagram. However, if a quantum particle is in an energy eigenstate, it has a definite energy and it is not appropriate to talk about different energies in different regions. 
Students often incorrectly asserted that a particle is in a bound state when it is in the classically allowed region and it is in a scattering state when it is in a classically forbidden region. Responses on other questions also indicate that the students did not realize that whether a state is a bound or a scattering state only depends on the energy of the particle compared to the potential energy at plus and minus infinity. \\

\textbf{Difficulties with graphing wavefunctions}:
In addition to upper-level studies, studies on introductory and intermediate level quantum mechanics have also found that students have difficulties in sketching the shape of a wavefunction~\cite{zollman2,brad,singhgrad}. Questions related to the shape of the wavefunction show that students may draw a qualitatively incorrect sketch even if their 
mathematical form of the wavefunction is correct, 
may draw wavefunctions with discontinuities or cusps,
or may confuse a scattering state wavefunction for a potential energy barrier problem with the wavefunction for a potential energy well problem.

In a multi-university study~\cite{singhgrad}, upper-level students were given the potential energy diagram for a finite square well. 
In part (a), they were asked to sketch the ground state wavefunction and in part (b) they had to sketch any one scattering state wavefunction.
In both cases, students were asked to comment on the shape of the wavefunction inside and outside the well. In part (a), students had to draw the ground
state wavefunction as a sinusoidal curve inside the well and with exponentially decaying tails in the classically forbidden regions. 
The wavefunction and its first derivative should be continuous everywhere and the wavefunction should be single valued. In part (b), they had to draw
a scattering state wavefunction showing oscillatory behavior in all regions, but because the potential energy is lower in the well, 
the wavelength is shorter in the well. For part (b), all graphs of functions that were oscillatory in both regions (regardless of the relative wavelengths 
or amplitudes in different regions) and showed the wavefunction and its first derivative as continuous
were considered correct. If the students drew the wavefunction correctly, their responses were considered correct even if they did not comment
on the shape of the wavefunction in the three regions. 

We note that this is one of the easiest questions involving the sketching of a wavefunction that upper-level students can be asked to do.
In response to this question, some students incorrectly drew as the ground state wavefunction for the infinite square well a curve that goes to zero in the classically forbidden
region. Others drew an oscillatory wavefunction in all three regions.
Many students drew either the
first excited state or a higher excited bound state with many oscillations in the well  and exponential decay outside. 
Some students incorrectly claimed that the particle is bound inside the well but free outside the well.
These types of student responses displayed confusion about what a ``bound state'' means and
whether the entire wavefunction is associated with the particle at a
given time or the parts of the wavefunction outside and inside
the well are associated with the particle at different times.
In part (b), some students 
drew a scattering state wavefunction that had an exponential decay in the well and
others drew wavefunctions with incorrect boundary conditions or that had
discontinuities or cusps in some locations. 
Although students were explicitly given a diagram of the potential energy well, 
responses suggest that some may be confusing the potential energy well with a potential energy barrier.
For example, some students plotted a wavefunction (without labeling the axes) which looked like 
a parabolic well with the entire curve drawn below the horizontal axis and claimed that the wavefunction must follow the sign of the potential energy.

\subsection{Difficulties with the time-dependence of a wavefunction}

The time-dependence of a quantum state or wave function is governed by the Time-Dependent Schr\"odinger Equation (TDSE)
\begin{eqnarray}
i \hbar  \frac{\partial \vert \Psi(t) \rangle}{\partial t}= \hat H \vert \Psi(t) \rangle \,\,\,\,\,\,\,or\,\,\,\,\,\,\,
i \hbar  \frac{\partial \Psi(x,t)}{\partial t}= \hat H \Psi(x,t) 
\end{eqnarray}
(where the second equation above is the TDSE for a particle confined in one spatial dimension in the position representation for which the Hamiltonian
$\hat H=\hat p^2/(2m)+\hat V$ in the position representation is $\hat H=  -\frac{\hbar^2}{2m} \frac{d^2}{dx^2}+ \hat V(x)$).

The TDSE shows that the time evolution of a wavefunction $\Psi(x,t)$ is governed by the Hamiltonian $\hat H$ of the system  and therefore the eigenstates of the Hamiltonian are special with respect to the time-evolution of a state.

When the Hamiltonian does not have an explicit time dependence, an equivalent way to represent the time evolution of the wavefunction is via
$\Psi(x,t)=e^{-i\hat H t/\hbar}\Psi(x,t=0)$.
In general, one can write $\Psi(0)=\Psi(x,t=0)=\sum_{n=1}^{\infty} C_n \phi_n$, where $\phi_n$ are the stationary
state wavefunctions for the given Hamiltonian with a discrete energy eigenvalue spectrum and $C_n=\langle \phi_n \vert \Psi \rangle$ are the expansion coefficients. Then, given any initial state of the system $\Psi(x,t=0)$, one can write
$\Psi(x,t)=e^{-i\hat H t/\hbar}\Psi(x,t=0)=\sum_{n=1}^{\infty} C_n e^{-iE_n t/\hbar} \phi_n $ where $E_n$ are the possible energies. It is clear from this form 
of $\Psi(x,t)$ which does not involve the Hamiltonian operator (but instead involves possible energies of the system, which are numbers) that only in the case in which the initial state is an energy eigenstate will the time-dependence of the system be trivial (because the wavefunction after a time $t$ will differ from the initial wavefunction only via an overall phase factor which does not alter measurement probabilities). For all other initial state wavefunctions, the time-dependence of the wavefunction will be non-trivial and, in general, the probabilities of measuring different observables will be time-dependent.

The following difficulties with the time-dependence of the wavefunction were commonly found via research.\\

\textbf{Incorrect belief that the Time-Independent Schr\"odinger Equation is the most fundamental equation in quantum mechanics}:
The most common difficulties with quantum dynamics are coupled with a focus on the Time-Independent Schr\"odinger Equation (TISE).
The time evolution of a wavefunction $\Psi(x,t)$ is governed by the Hamiltonian $\hat H$ of the system via
the TDSE, and thus, the TDSE is considered the most fundamental equation of quantum mechanics. Since there are no dynamics in the TISE, focusing on the TISE as the most fundamental equation in quantum mechanics leads to difficulties.
For example, in Ref.~\cite{singhgrad}, students were asked to write down the most fundamental equation of quantum mechanics.
Approximately one-third of the students provided a correct response while half of them claimed
that the TISE is the most fundamental equation of quantum mechanics.
It is true that if the potential energy is time-independent, one can use separation of variables
to obtain the TISE, which is an eigenvalue equation for the Hamiltonian. The eigenstates of $\hat H$ obtained by solving the TISE are
stationary states which form a complete set of states so that any general wavefunction
can be written as a linear superposition of the stationary states. However, overemphasis on the TISE and de-emphasis on the TDSE in quantum mechanics courses result in many students struggling with the time-dependence of a wavefunction.\\

\textbf{Incorrect belief that the time-evolution of a wavefunction is always via an overall phase factor of the type ``$e^{-iE t/\hbar}$'' }:
Due to excessive focus on the TISE and stationary state wavefunctions, many students claim that given any $\Psi(x,t=0)$, one can find the wavefunction after time $t$ using ``$\Psi(x,t)=e^{-iE t/\hbar}\Psi(x,t=0)$'' where $E$ is a constant. For example, in Ref.~\cite{singhgrad}, students from seven universities were given a linear superposition of the ground and first excited state wavefunction as the initial wavefunction ($\Psi(x,t=0)=\sqrt{2/7} \phi_1(x) +\sqrt{5/7} \phi_2(x) $) for an electron in a one-dimensional infinite square well and asked to find the wavefunction $\Psi(x,t)$ after a time $t$. 

Instead of the correct response,
$\Psi(x,t)=\sqrt{2/7} \phi_1(x) e^{-iE_1t/\hbar}+\sqrt{5/7} \phi_2(x) e^{-iE_2t/\hbar}$, in which the ground state wave function is $\phi_1(x)$ and the first excited state wavefunction is $\phi_2(x)$, approximately one-third of students wrote common phase factors for both terms, e.g.,
``$\Psi(x,t)=\Psi(x,0)e^{-iEt/\hbar}$.'' Interviews suggested that these students were having difficulty differentiating
between the time-dependence of stationary and non-stationary state wavefunctions. Students struggled with the fact that since the Hamiltonian operator governs the time-development of the system, the time-dependence of a stationary state wavefunction is via a simple phase factor but non-stationary state wavefunctions, in general,
have a non-trivial time-dependence because each term in a linear superposition of stationary states evolves via a different phase factor.
Apart from using ``$e^{-iEt/\hbar}$'' as the common phase factor, other common choices include ``$e^{-i \omega t}$'', ``$e^{-i \hbar t}$,''
``$e^{-it}$,'' ``$e^{-ixt}$,'' ``$e^{-ikt}$,'' etc.

In the context of a non-stationary state wavefunction which is not explicitly written as a linear superposition of stationary states, similar difficulties are observed. For example, in a study involving ten different universities, students were asked to select the correct probability density after a time $t$ for an initial normalized wavefunction $Asin^5(\pi x/a)$ in an infinite square well potential. In response to this question, half of the students incorrectly claimed that the probability density is time-independent because of the overall time-dependent phase factor in the wavefunction which cancels out in probability density ~\cite{zhuqms1}.\\

\textbf{Inability to differentiate between $e^{-i\hat H t/\hbar}$ and  $e^{-iE t/\hbar}$ }:
In Ref.~\cite{singhgrad}, in response to the question asking for $\Psi(x,t)$ given an initial state which is a linear superposition of the ground and first excited states, $\Psi(x,t=0)=\sqrt{2/7} \phi_1(x) +\sqrt{5/7} \phi_2(x) $, some students wrote incorrect intermediate steps; e.g.,
``$\Psi(x,t)=\Psi(x,0)e^{-iEt/\hbar}=\sqrt{2/7} \phi_1(x) e^{-iE_1t/\hbar}+\sqrt{5/7} \phi_2(x) e^{-iE_2t/\hbar}$.''
Probing during the individual interviews  showed that these students had difficulty differentiating between the Hamiltonian
operator and its eigenvalue and incorrectly used ``$\hat H=E$'' where $E$ is a number instead of
$\Psi(x,t)=e^{-i\hat H t/\hbar}\Psi(x,0)=\sqrt{2/7} \phi_1(x) e^{-iE_1t/\hbar}+\sqrt{5/7} \phi_2(x) e^{-iE_2t/\hbar}$ where
the Hamiltonian $\hat H$ acting on the stationary states gives the corresponding energies~\cite{improve}. The inability to differentiate between the Hamiltonian operator and energy can reinforce the difficulty that all wavefunctions evolve via an overall phase factor of the type ``$e^{-iE t/\hbar}$.''\\

\textbf{Incorrect belief that for a time-independent Hamiltonian, the wavefunction does not depend on time}:
Some students claimed that $\Psi(x,t)$  should not have {\it any} time dependence whatsoever if the Hamiltonian does not have an explicit time-dependence. For example, in response to the question about the time-dependence of the wavefunction given an initial state which is a linear superposition of the ground and first excited states ($\Psi(x,t=0)=\sqrt{2/7} \phi_1(x) +\sqrt{5/7} \phi_2(x) $), some students claimed that there is no time dependence and typically justified their answer by pointing to the TISE and adding that the Hamiltonian is not time-dependent so there cannot be any time-dependence to the wavefunction~\cite{singhgrad}.\\

\textbf{Incorrect belief that the time-dependence of a wavefunction is represented by a real exponential function}:
Some students claimed that the time dependence of a wavefunction, e.g., an initial wavefunction $\Psi(x,t=0)=\sqrt{2/7} \phi_1(x) +\sqrt{5/7} \phi_2(x)$, is a decaying exponential,
e.g., of the type ``$\Psi(x,0)e^{-xt}$,'' ``$\Psi(x,0)e^{-Et}$,''  ``$\Psi(x,0)e^{-ct}$,'' ``$\Psi(x,0)e^{-t}$,'' etc. During the interviews,
some of these students explained their choices by insisting that the wavefunction must decay with time because
``this is what happens for all physical systems''~\cite{singhgrad}.

\subsection{Difficulties in distinguishing between three-dimensional Euclidian space and Hilbert space}

In quantum theory, it is necessary to interpret the outcomes of real experiments performed in real space by
making a connection with an abstract Hilbert space (state space) in which the state of the system or wavefunction lies. The
physical observables that are measured in the laboratory correspond to Hermitian operators in the Hilbert space in which the state of the system lies. Knowing the initial wavefunction and the Hamiltonian of the system allows
one to determine the time-evolution of the wavefunction unambiguously and the measurement postulate can be used to 
determine the possible outcomes of individual measurements and ensemble averages (expectation values) at a given time.
Research suggests that students have the following types of difficulties about these issues:\\

\textbf{Difficulties in distinguishing between vectors in real space and Hilbert space}:
It is difficult for students to distinguish between vectors in real space and Hilbert space. For example, 
$S_x$, $S_y$ and $S_z$ denote the orthogonal components of the spin angular momentum vector of an electron in three dimensions, each 
of which is a physical observable that can be measured in the laboratory. However, the Hilbert space corresponding to the
spin degree of freedom for a spin-1/2 particle is two-dimensional (2D). In this Hilbert space, $\hat S_x$, $\hat S_y$ and $\hat S_z$
are operators whose eigenstates span the 2D space~\cite{sga}.
The eigenstates of $\hat S_x$ are vectors which span the 2D space and are orthogonal to each other
(but not orthogonal to the eigenstates of $\hat S_y$ or $\hat S_z$). 
Also, $\hat S_x$, $\hat S_y$ and $\hat S_z$ are operators and not orthogonal components of a vector in 2D space. If the electron is in a magnetic field with a gradient in the $z$-direction in the laboratory (real space) as in a Stern-Gerlach experiment, the magnetic field is a vector field
in three-dimensional (3D) space and not in 2D Hilbert space. It does not make sense to compare vectors in 3D space with
vectors in the 2D space as in statements such as ``the magnetic field gradient is perpendicular to the eigenstates of $\hat S_x$.''
However, these distinctions are difficult for students to make and such difficulties are common as discussed in Refs.~\cite{singhgrad,singhtransfer}.

For example, in a multi-university study in Ref.~\cite{singhgrad}, these types of difficulties were found in student responses to a question
related to the Stern-Gerlach experiment. 
Students were told that the notation $\vert \uparrow_z \rangle$ and $\vert \downarrow_z \rangle$ represents the orthonormal eigenstates of  $\hat S_z$ (the $z$ component of the spin angular momentum) of a spin-1/2 particle.
In the situation in the question, a beam of electrons propagating along the $y$-direction (into the page) in spin state 
$\vert \uparrow_z \rangle$ is sent through an apparatus with a horizontal magnetic field gradient in the $-x$-direction. 
Students were asked to sketch the electron cloud pattern they expect to see on a distant phosphor screen 
in the $x$-$z$ plane and explain their reasoning. This question is challenging  because students have to realize that the eigenstate of $\hat S_z$, $\vert \uparrow_z \rangle$, can be written as a linear superposition of the eigenstates of $\hat S_x$, that is, 
$\vert \uparrow_z \rangle=(\vert \uparrow_x \rangle + \vert \downarrow_x \rangle) /\sqrt{2}$. 
Therefore, the magnetic field gradient in the $-x$-direction will 
split the beam along the $x$-direction corresponding to the electron spin components 
in $\vert \uparrow_x \rangle$ and $\vert \downarrow_x \rangle$ states and cause two spots on the phosphor screen.
The most common difficulty was assuming that 
there should not be any splitting since the magnetic field gradient (in the $-x$-direction) and
the spin state (an eigenstate of $\hat S_z$) are orthogonal to each other. It can be inferred from the responses that 
students incorrectly relate the direction of the magnetic field in real space 
with the ``direction'' of the state vectors in Hilbert space. \\

\textbf{Difficulties in distinguishing between the dimension of physical space and Hilbert space}:
The dimension of a Hilbert space is equal to the number of linearly
independent basis vectors. The linearly independent 
eigenstates of an operator corresponding to an observable may be used as basis vectors. For example, for a particle in a one-dimensional
(1D) infinite square well, the infinitely many energy eigenstates $\left|
{\phi _n } \right\rangle $ of the Hamiltonian operator form a complete set
of basis vectors for the infinite-dimensional Hilbert space.
However, students have great difficulty in distinguishing between the dimensions of the Hilbert space and the dimensions of the physical space. For example, in a multiple choice question about the dimensionality of the Hilbert space for a 1D infinite square well~\cite{angular2}, less than half of the students provided the correct answer. The rest of the students claimed that the Hilbert space for this system is 1D
and that the position eigenstates and energy eigenstates of the system form a basis 
for the one-dimensional Hilbert space (students did not realize that they were making contradictory statements because there is not only one but infinitely many energy eigenstates or position eigenstates) for this quantum system.

\subsection{Difficulties with measurements and expectation values}

If the wavefunction is known right before a measurement, quantum theory only provides the probability of measurement outcomes when an observable is measured. After the measurement, the state of the system collapses into an eigenstate of the operator corresponding to the observable measured. The expectation value of an observable $Q$ in a state is the average value of a large number of measurements of $Q$ on identically prepared systems. Since measurement outcomes are probabilistic if the state is not in an eigenstate of $\hat Q$, an ensemble average is useful because it is deterministic for a given quantum state of a system.
Research suggests that students have great difficulties with quantum measurement~\cite{singh2001,singhtoday,singhgrad,gire1,measure01,measure02,measure1,measure2}.\\

\textbf{Difficulties with the probability of a particular outcome of a measurement}:
When calculating the probability of obtaining a certain value in the measurement of a physical observable, students often incorrectly claim that the operator corresponding to that observable must be explicitly involved in the expression~\cite{brad}. For example, in a multi-university multiple-choice survey~\cite{zhuqms1}, students were  asked to suppose that a particle in a one-dimensional infinite square well is in the ground state with wavefunction $\phi_1(x)$ and they had to find the probability that the particle will be found in a narrow range between $x$ and $x+dx$. In response to this question, approximately one-third of the students chose the distractor ``$\int_{x}^{x+dx} x \vert \phi_1(x) \vert^2 dx$'' as the probability of finding the particle in the region between $x$ and $x+dx$. They did not recognize that $\vert \phi_1(x) \vert^2 dx$ is the probability of finding the particle between $x$ and $x+dx$. In another question on the same survey~\cite{zhuqms1}, students were given a non-stationary state wavefunction $\Psi(x,0)=Ax(a-x)$ for an infinite square well and they were asked to select the correct expression, $\vert \int_0^a \phi_n^\star(x) \Psi(x,0) dx \vert^2$, for the probability of measuring energy $E_n$. Less than half of the students provided the correct response and  one-third of the students incorrectly claimed that ``$\vert \int_0^a \phi_n^\star(x) \hat H \Psi(x,0) dx \vert^2$'' is the probability of measuring the energy. Students often did not realize that the required information about the energy measurement is obtained by projecting the state of the system along the energy eigenstate (multiplication of the wavefunction by $\phi_n^\star(x)$ before integrating). \\

\textbf{Difficulties with the possible outcomes of a measurement}:
According to the Copenhagen interpretation, the measurement of a physical observable instantaneously collapses the state to an eigenstate of the corresponding operator and the corresponding eigenvalue is measured. In Ref.~\cite{zhuqms1}, some questions on the survey investigated students' understanding of the energy measurement outcomes, e.g., for a superposition of two stationary states $\Psi(x,0)=\sqrt{2/7} \phi_1(x)+\sqrt{5/7} \phi_2(x)$ of a 1D infinite square well. 
The only possible results of the energy measurement are the ground state energy $E_1$ and the first excited state energy $E_2$. When the energy $E_2$ is obtained, the wavefunction of the system collapses to $\phi_2(x)$ and remains there. However,  students often incorrectly claimed that the normalized collapsed wavefunction should be ``$\sqrt{5/7} \phi_2(x)$,'' which has an incorrect normalization. Also, one-third incorrectly claimed that the wavefunction would collapse first but finally evolve back to the initial state. Other students did not realize that the wavefunction would collapse and claimed that the system will remain in the initial state even after the measurement. \\

\textbf{Difficulties in distinguishing between eigenstates of operators corresponding to different observables}:
A very common difficulty is assuming that eigenstates of operators corresponding to all physical observables are the same~\cite{singh2001,singhgrad,singhtoday}.
The measurement of a physical observable collapses the wavefunction of a quantum system into an eigenstate of the corresponding operator. Many students had difficulties in distinguishing between energy eigenstates and the eigenstates of other physical observables. In a multi-university survey~\cite{zhuqms1}, half of the students claimed that the stationary states refer to the eigenstates of any operator corresponding to a physical observable because they had difficulty in differentiating between the related concepts of stationary states and eigenstates of other observables. Many students claimed that in an isolated system, if a particle is in a position eigenstate (has a definite value of position) at time $t=0$, the position of the particle is well-defined at all times $t>0$. Students did not relate the stationary state with the special nature of the time evolution in that state (the state evolves via an overall phase factor so that the measurement probabilities for observables do not depend on time). 
In another study~\cite{singhgrad}, some students claimed that the wavefunction will become peaked about a certain value of position and drew a delta function in position when asked to draw the wavefunction after an energy measurement.\\

\textbf{Confusion between the probability of measuring position and the expectation value of position}:
Born's probabilistic interpretation of the wavefunction can also be confusing for students.
In a multi-university investigation~\cite{singhgrad}, students were told that for an electron in a 1D infinite square well, immediately after an energy measurement which yields the first excited state energy 
$4\pi^2 \hbar^2/(2ma^2)$, the position of the electron is measured.  They were asked to qualitatively describe the possible values of position one can measure and the probability of measuring them. The correct answer involves noting that it is possible to measure position values between $x=0$ and $x=a$ (except at $x=0$, $a/2$, and $a$ where the wavefunction is zero), and according to Born's interpretation, ``$\vert \phi_2(x) \vert^2 dx$'' gives the probability of finding the particle between $x$ and $x+dx$ if $\phi_2(x)$ is the first excited state. Less than half of the students provided the correct response. Many students tried to find the expectation value of position $\langle x \rangle$ instead of the probability of finding the electron
at a given position. They wrote the expectation value of position in terms of an integral involving the wavefunction. Others explicitly wrote that ``$\mbox{Probability}=\!(2/a)\!\int_0^a x \sin^2(2\pi x/a) dx$'' and claimed that instead of $\langle x \rangle$ they were calculating the probability of measuring the position of the electron.
Some students justified their response by incorrectly claiming that $\vert \Psi \vert^2$
gives the probability of the wavefunction being at a given position and if you multiply it by $x$ you get the probability of measuring the position $x$.\\ 

\textbf{Difficulties with measuring energy after position measurement}:
In a multi-university investigation~\cite{zhuqms1}, one question examined students' understanding of consecutive quantum measurements, e.g., measuring the energy of a quantum system immediately after a position measurement. 
For a one-dimensional infinite square well with an initial state which is a superposition of the ground and first excited states, the position measurement will collapse the wavefunction of the system to a delta function which is a superposition of infinitely many energy eigenfunctions. Therefore, one can obtain higher order energy values ($n>2$) for the energy measurement after the position measurement. However, less than one-third of the students correctly answered the question and realized that the state of the system changed after the position measurement. More than one-third mistakenly claimed that they can only obtain either energy $E_1$ or $E_2$, which correspond to the initial state before the position measurement.\\

\textbf{Difficulties with measuring position after energy measurement}:
In another multi-university study~\cite{singhgrad}, one question asked students to qualitatively describe the possible values of the position of an electron that one can measure if the position measurement follows an energy measurement which yields the first excited state energy. In response to this question, some students tried to use the generalized uncertainty principle between energy and position or between position and momentum,
but most of their arguments led to incorrect inferences. According to the generalized uncertainty principle, if     $\sigma_A$ and $\sigma_B$ are the standard deviations in the measurement of two observables $A$ and $B$,  respectively, in a state $\vert \Psi \rangle$,  and $[\hat A, \hat B]$ is the commutator of the operators corresponding to $A$ and $B$, respectively, then $ \sigma_A^2 \sigma_B^2 \ge ({\langle \Psi\vert [\hat A,\hat B]\vert \Psi \rangle}/(2 i) )^2$.

Although the generalized uncertainty principle implies that position and energy are indeed incompatible observables since their corresponding operators do not commute, students often made incorrect inferences to answer the question posed. For example, several students noted that because the energy is well-defined immediately after the measurement of energy, the uncertainty in position must be infinite according to the uncertainty principle. Some students even went on to argue that the probability of measuring the particle's position is the same everywhere using the generalized uncertainty principle. Others restricted themselves only to the inside of the well and noted that the uncertainty principle says that the probability of finding the particle is the same everywhere inside the well and for each value of position inside the well this constant probability is ``$1/a$.''
These students typically claimed that the particle must be between $x=0$ and $x=a$ but by knowing the exact energy, we can know nothing about position so the probable position is spread uniformly within in $0<x<a$ region.
Some students thought that the most probable values of position were the only possible values of the position that can be measured.
The following statement was made by a student who thought that it may not be possible to measure the position after measuring the energy: 
``Can you even do that? Doesn't making a measurement change the system in a manner that makes another measurement invalid?''
The fact that the student felt that making a measurement of one observable 
can make the immediate measurement of another
observable invalid sheds light on the student's epistemology about quantum theory.\\

\textbf{Difficulties with interpreting the expectation value as an ensemble average}:
Many students have difficulty in interpreting the expectation value as an ensemble average. For example, in a multi-university survey~\cite{singhgrad}, students were given the wavefunction of an electron in a one-dimensional infinite square well as a particular linear superposition of ground and first excited states ($\Psi(x,t=0)=\sqrt{2/7} \phi_1(x) +\sqrt{5/7} \phi_2(x)$). They were asked to write down the possible outcomes of energy measurement and their probabilities in part (I) and then calculate the expectation value of the energy in state $\Psi(x,t)$ in part (II). 

In part (I), two-thirds of the students correctly stated that the only possible values of the energy in state 
$\Psi(x,0)$ are $E_1$ and $E_2$ and their respective probabilities are $2/7$ and $5/7$.  But only slightly more than one-third provided the correct response for part (II).
 The discrepancy in percentages is due to the fact that many students who could calculate probabilities for the
possible outcomes of energy measurement were unable to use that information
to determine the expectation value of the energy. Since the expectation value of the energy is
time-independent, if $\Psi(x,t)=C_1(t) \phi_1(x)+C_2(t) \phi_2(x)$, 
then the expectation value of the energy in this state is 
$\langle E \rangle=P_1 E_1+P_2 E_2=\vert C_1(t)\vert^2 E_1+ \vert C_2(t)\vert^2 E_2=(2/7)E_1 +(5/7)E_2$,
where $P_i=\vert C_i(t)\vert^2$ is the probability of measuring the energy $E_i$ at time $t$. However, many students who answered part (II) correctly calculated $\langle E\rangle $ by ``brute-force'': first writing $\langle E\rangle
=\!\int_{-\infty }^{+\infty }\Psi^{\ast}\hat{H}\Psi dx$, expressing $\Psi(x,t)$ 
in terms of the linear superposition of two energy eigenstates, then acting with the operator $\hat{H}$ on the eigenstates, and finally using orthogonality to obtain the answer. Some got lost early in this process. Others did not remember some steps, for example, taking the complex conjugate of the wavefunction, using the orthogonality of stationary states, or recognizing the proper limits of the integral. The interviews revealed that 
many students did not know or recall the interpretation of expectation value as an ensemble average and did not recognize that expectation values could be calculated more simply in this case by taking advantage of their answer to part (I).\\

\textbf{Confusion between individual measurements vs. expectation value}:
In response to the question discussed in the preceding section~\cite{singhgrad}, some students who were asked about possible values of an energy measurement and their probabilities in a particular superposition of the ground and first excited state wavefunctions became confused between individual measurements of the energy and its expectation value. Almost none of them calculated the correct expectation value of the energy. \\

\textbf{Incorrect assumption that all energies are possible when the state is in a superposition of only the ground and first excited states}:
In response to the question discussed in the preceding section~\cite{singhgrad}, another mistake students made was assuming that all allowed energies
for the infinite square well were possible if the measurement of energy was performed when the system was in state $\sqrt{2/7} \phi_1(x) +\sqrt{5/7} \phi_2(x)$ and that the ground state energy is the most probable measurement outcome because it is the lowest energy state.\\

\textbf{Difficulties with time development of the wavefunction after measurement of an observable}:
In a multi-university investigation~\cite{zhuqms1}, students were told that a
 measurement of the position of the particle is performed when it is in the first excited state of a one-dimensional finite square well and were asked about the time development of the wavefunction after the measurement. 
More than one-third of the students incorrectly claimed that the wavefunction of the system after a position measurement will go back to the first excited state (which was the state before the measurement was performed) after a long time. Other students who provided incorrect responses often claimed that the wavefunction was stuck in the collapsed state after the measurement. In one-on-one interview situations, when these students were told explicitly that their initial responses were not correct and they should think about what quantum mechanics predicts about what should happen to the wavefunction after a long time, students often switch from stating ``it goes back to the original wavefunction before measurement'' to ``it remains stuck in the collapsed state'' and vice versa. When students were told that neither of the possibilities is correct and that they should think about what quantum mechanics actually predicts, some of them explicitly asked the interviewer how there can be any other possibility. Thus, students have great difficulty with this three-part problem in which 1) the particle is initially in the first excited state of a 1D infinite square well, 2) a measurement of position collapses the wavefunction of the particle at the instant the measurement is performed, and 3) the wavefunction evolves again according to the TDSE. Connecting the different parts of this situation is extremely challenging for advanced students. 

In the same multi-university investigation~\cite{zhuqms1}, students were told that the wave function for the system is $\sqrt{2/7} \phi_1(x)+\sqrt{5/7} \phi_2(x)$ when a measurement of energy is performed.
They were asked about the wavefunction a long time after the measurement if the energy measurement yields $4\pi^2 \hbar^2/(2ma^2)$. Less than half of the students provided the correct response and 
an equal percentage claimed that a long time after the measurement, the system will be in the original superposition state $\sqrt{2/7} \phi_1(x)+\sqrt{5/7} \phi_2(x)$.

In response to a similar question~\cite{singhgrad}, 
some students claimed that the answer to the question depends upon how much time you wait after the measurement. They claimed that at the instant you measure the energy,
the wavefunction will be $\phi_2$, but if you wait long enough it will go back to the state before the measurement. The notion that the
system must go back to the original state before the measurement was sometimes deep-rooted. For example, when an interviewer said to a student that it was not clear why that would
be the case, the student said, ``The collapse of the wavefunction is temporary \ldots\ Something has to happen to the wavefunction for
you to be able to measure energy or position, but after the measurement the wavefunction must go back to what it {\it actually} 
(student's emphasis) is supposed to be.'' When probed further, the student continued, ``I remember that if you measure 
position you will get a delta function, but it will stay that way only if you do repeated measurements \ldots\ if you let it 
evolve it will go back to the previous state (before the measurement).''
Some students confused the measurement of energy with the measurement of position and drew a delta function for what the wavefunction will look like after the energy measurement. They  claimed that the wavefunction will become very peaked about a given position after the energy measurement. As for the time evolution after that, students with these types of responses either incorrectly claimed that the system would be stuck in that peaked state or will evolve back to the original state of the system.\\

\textbf{Incorrect belief that the Hamiltonian acting on a state represents energy measurement}:
In a multi-university investigation~\cite{singhtoday}, students were asked to argue whether or not ``$\hat H \Psi=E \Psi$'' is always true for any possible $\Psi$ of the system. Many students incorrectly claimed that any statement involving a Hamiltonian operator acting on a state is a statement about the measurement of energy. Some of the students who incorrectly claimed that ``$\hat H \Psi=E \Psi$'' is a statement about energy measurement agreed that the statement ``$\hat H \Psi=E \Psi$'' is always true, while others disagreed. Those who disagreed often claimed that ``$\hat H \Psi=E_n \phi_n$,''  because as soon as $\hat H$ acts on $\Psi$, the wavefunction will
collapse into one of the energy eigenstates $\phi_n$ and the corresponding energy $E_n$ will be obtained. For example, one student stated:  ``Agree. $\hat H$ is the operator for an energy measurement. Once this measurement takes place, the specific value $E$ of the energy will be known.''
The interviews and written answers suggest that these students thought that 
the measurement of a physical observable in a particular state 
is achieved by acting with the corresponding operator on the state.
These incorrect notions are overgeneralizations of the fact that the Hamiltonian operator corresponds to energy and 
after the measurement of energy, the system is in a stationary state so $\hat H \phi_n=E_n \phi_n$. This example illustrates the difficulty students have in relating the formalism of quantum mechanics to the measurement of a physical observable. 

In other investigations~\cite{analogous,my1}, over half of the students claimed that 
either ``$\hat Q \vert \Psi \rangle =q_n \vert \Psi \rangle $,'' ``$\hat Q \vert \Psi \rangle =q_n \vert \psi_n \rangle $,'' or both expressions are correct for a system in a state $\vert \Psi \rangle$ which is not an eigenstate of $\hat Q$. Neither of the aforementioned expressions is correct in terms of linear algebra. The response rates are very similar when the question is asked explicitly about the Hamiltonian operator. Thus, in this case, when ``$\hat H \vert \Psi \rangle =E_n \vert \phi_n \rangle $'' is explicitly brought to students' attention, more students are primed to select ``$\hat H \vert \Psi \rangle =E_n \vert \phi_n \rangle $'' as true compared to the case when they are asked the question in an open ended format (i.e., when they are asked if ``$\hat H \Psi=E \Psi$'' is always true for all possible wavefunctions)~\cite{singhtoday}. This difference in the percentages of students who select a particular incorrect response depending on whether some common difficulty was explicitly mentioned to prime students was discussed at the beginning of this section. This type of context dependence of responses is a sign of the fact that students do not have a robust knowledge structure of quantum mechanics.\\

\textbf{Incorrect belief that ``$\hat Q \Psi=\lambda \Psi$'' is true for all possible $\Psi$ of the system for any physical observable $Q$}:
In general, $\hat Q \Psi \ne \lambda \Psi$ unless $\Psi$
is an eigenstate of $\hat Q$ with eigenvalue $\lambda$. A generic state $\Psi$ can be represented as
$\Psi=\sum_{n=1}^{\infty} D_n \psi_n$, where $\psi_n$ are the eigenstates of $\hat Q$
and $D_n=\langle \psi_n \vert \Psi \rangle$. Then, $\hat Q \Psi=\sum_{n=1}^{\infty} D_n \lambda_n \psi_n$ (for an
observable with a discrete eigenvalue spectrum).
In Ref.~\cite{singhgrad}, individual interviews suggest that some students thought that 
if an operator $\hat Q$ corresponding to a physical observable $Q$ acts on any state $\Psi$, it will
yield the corresponding eigenvalue $\lambda$ and the same state back, that is, ``$\hat Q \Psi=\lambda \Psi$''~\cite{singhgrad}.
Some of these students were overgeneralizing their incorrect ``$\hat H \Psi=E \Psi$'' reasoning and attributing
``$\hat Q \Psi=\lambda \Psi$'' to the measurement of an observable $Q$.\\

\noindent
\textbf{Incorrect belief that an operator acting on a state represents a measurement of the corresponding observable}:
In Ref.~\cite{singhgrad}, some students overgeneralized their incorrect notion that ``$\hat H \Psi=E_n \phi_n$'' to conclude that
``$\hat Q \Psi=\lambda_n \psi_n$'' must be true. They claimed that this equation is a statement about the measurement of $Q$ which collapses the wavefunction into an eigenstate of $\hat Q$ corresponding to the eigenvalue $\lambda_n$ measured~\cite{singhgrad}.

\subsection{Difficulties with the time-dependence of expectation values}

Generally, the expectation value of an observable $Q$ evolves in time because the state of the system evolves in time in the Schr\"odinger formalism.  
If an operator \op{Q} corresponding to an observable $Q$ has no explicit time dependence (assumed throughout), taking the time derivative of the states in the expectation value and making use of the TDSE where appropriate yields Ehrenfest's theorem: $d\braket{Q(t)}/dt = \braket{\Psi(t) | [\op{Q},\op{H}] | \Psi(t)} / {i \hbar} $.
Two major results can be deduced from this theorem:
(1)  The expectation value of an operator that commutes with the Hamiltonian 
 is time-independent regardless of the initial state; and
(2) If the system is initially in an energy eigenstate, the expectation value of
any operator $\hat Q$ will be time-independent.
The following student difficulties were commonly found via research~\cite{singh2001,marshmantime,ben2014}:\\

\textbf{Difficulties in recognizing the relevance of the commutator of an operator corresponding to an observable and the Hamiltonian \label{H}}:
A consequence of Ehrenfest's Theorem is that if an operator \op{Q} corresponding to an observable $Q$ commutes with the Hamiltonian, the time derivative of $\braket{Q}$ is zero, regardless of the state.  However, approximately half of students~\cite{marshmantime} did not realize that since the Hamiltonian governs the time-evolution of the system, any operator \op{Q} that commutes with it must correspond to an observable which is a constant of motion and its expectation value must be time-independent.\\

\textbf{Difficulties in recognizing the special properties of stationary states}: 
In the context of Larmor precession, if the magnetic field is along the $z$-axis, all expectation values are time independent if the initial state is an eigenstate of \op{S_z} because it is a stationary state.   However, half of students~\cite{marshmantime} incorrectly stated that $\braket{S_x}$ and $\braket{S_y}$ depend on time in this case.  One common difficulty includes reasoning such as ``since the system is not in an eigenstate of \op{S_x}, the associated expectation value must be time dependent,'' even in a stationary state.  Another very common difficulty is reasoning such as ``since \op{S_x} does not commute with \op{H}, its expectation value must depend on time,'' even in a stationary state.\\

\textbf{Difficulties in distinguishing between stationary states and eigenstates of operators corresponding to observables other than energy}:
Any operator corresponding to an observable has an associated set of eigenstates, but only eigenstates of the Hamiltonian are stationary states because the Hamiltonian plays a central role in the time-evolution of the state.  However, many students were unable to differentiate between these concepts. For example, for Larmor precession with the magnetic field in the $z$-direction, half of the students~\cite{marshmantime} claimed that if a system is initially in an eigenstate of \op{S_x} or \op{S_y}, the system will remain in that eigenstate.  A related common difficulty is exemplified by the following comment from a student: ``if a system is initially in an eigenstate of \op{S_x}, then only the expectation value of $S_x$ will not depend on time.''\\  

\textbf{These difficulties related to the time-dependence of expectation values were often due to the following types of overgeneralizations or confusions:}
	
	\begin{itemize}
		\item An eigenstate of any operator is a stationary state.
	\item If the system is initially in an eigenstate of $\hat Q$, then the expectation value of that operator is time independent.
	\item If the system is initially in an eigenstate of any operator $\hat Q$, then the expectation value of another operator $\hat Q^\prime$ will be time independent if $[\hat Q,\hat Q^\prime]=0$.
	\item If the system is in an eigenstate of any operator $\hat Q$, then it remains in the eigenstate of $\hat Q$ forever unless an external perturbation is applied.
	\item The statement ``the time dependent exponential factors cancel out in the expectation value'' is synonymous with the statement ``the system does not evolve in any eigenstate.''
	\item The expectation value of an operator in an energy eigenstate may depend upon time.
	\item If the expectation value of an operator $\hat Q$ is zero in some initial state, the expectation value cannot have any time dependence.
	\item Individual terms in a time-independent Hamiltonian involving a magnetic field can cause transitions from one energy eigenstate to another. Therefore, being in a stationary state of a harmonic oscillator potential energy system is different from being in a stationary state of a system in which an electron is at rest in a uniform magnetic field. In the latter case, the expectation values will depend on time in a stationary state but not for the former (because there is no field to cause a transition).
	\item Time evolution of a state cannot change the probability of obtaining a particular outcome when any observable is measured regardless of the
	initial state because the time evolution operator is of the form $e^{-i \hat H t/\hbar}$, so time-dependent terms cancel out. Also, since $\vert \Psi (t) \rangle=e^{-i\hat Ht} \vert \Psi (t=0) \rangle $, the expectation value of any observable $Q$ in a generic state $\langle \Psi (t) \vert  \hat Q \vert \Psi (t) \rangle$ is time-independent.
	\end{itemize}
	
\subsection{Difficulties with the addition of angular momentum}

For a system consisting of two spin-1/2 particles, the Hilbert space is four dimensional. 
There are two common ways to represent the basis vectors for the product space. Since the spin quantum
numbers $s_1 =1/2$ and $s_2 =1/2$ are fixed, we can use the ``uncoupled
representation'' and express the orthonormal basis vectors for the product
space as $\left| {s_1, m_1 } \right\rangle \otimes \left| {s_2, m_2 }
\right\rangle =\left| {m_1 } \right\rangle \otimes \left| {m_2 }
\right\rangle $. In this uncoupled representation, the
operators related to each particle (subspace) act on their own states, e.g.,
$\hat {S}_{1z} \left| {1/2} \right\rangle _1 \otimes \left| {-1/2}
\right\rangle _2 =\frac{\hbar }{2}\left| {1/2} \right\rangle _1 \otimes
\left| {-1/2} \right\rangle _2 $ and $\hat {S}_{2z} \left| {1/2}
\right\rangle _1 \otimes \left| {-1/2} \right\rangle _2 =-\frac{\hbar
}{2}\left| {1/2} \right\rangle _1 \otimes \left| {-1/2} \right\rangle _2 $.
On the other hand, we can use the ``coupled representation'' and find the
total spin quantum number for the system of two particles together. The
total spin quantum number for the two spin-1/2 particle system, $s$, is either
$1/2+1/2=1$ or $1/2-1/2=0$. When the total spin quantum number $s$ is 1, the
quantum number $m_s $ for the $z$-component of the total spin, $S_z $, can be
1, 0, and $-$1. When the total spin is 0, $m_s $ can only be 0. Therefore, the
basis vectors of the system in the coupled representation are $\left|
{s=1,m_s =1} \right\rangle $, $\left| {s=1,m_s =0} \right\rangle $, $\left|
{s=1,m_s =-1} \right\rangle $ and $\left| {s=0,m_s =0} \right\rangle $. In
the coupled representation, the state of a two-spin system is not a simple
product of the states of each individual spin although we can write each
coupled state as a linear superposition of a complete set of uncoupled states. 
The following is a summary of the common difficulties students have with the addition of angular momentum~\cite{angular1,angular2}.\\

\textbf{Difficulties with the dimension of a Hilbert space in product space}:
Students often incorrectly assumed that the dimension
$D$ of a product space consisting of two subspaces of dimensions $D_1 $ and
$D_2 $ is $D=D_1 +D_2 $, stating that this was true for the following reasons: 
(1) We are ``adding'' angular momentum; and
(2) For two spin-one-half systems, the dimension is four which is
both $2\times 2$ and $2+2$.
\\

\textbf{Difficulties in identifying different basis vectors for the product space}:
Students often displayed the following difficulties in identifying different basis vectors for the product space:
(1) Some had difficulties with choosing a convenient basis to
represent an operator as an $N\times N$ matrix in an
N-dimensional product space;
and (2) Some incorrectly claimed that if the operator
matrix is diagonal in one representation, it must also be diagonal in
another representation.
\\

\textbf{Difficulties in constructing an operator matrix in the product space and realizing that the matrix corresponding to an operator could be very different in a different basis}:
Students displayed the following difficulties in constructing an operator matrix in the product space:
(1) Mistakenly adding the operators in different
Hilbert spaces algebraically to construct the operator for the product space
as if they act in the same Hilbert space; 
(2) Incorrectly claiming that the dimension of
the operator matrix depends on the choice of basis vectors and it is lower for the uncoupled representation compared to the coupled representation;
(3) Incorrectly assuming, e.g., that if $\hat S_z=\hat
{S}_{1z} +\hat {S}_{2z} $ is diagonal in the coupled representation, then $\hat
{S}_{1z} +\frac{1}{2}\hat {S}_{2z} $ must also be diagonal in that representation;
(4) Incorrectly assuming, e.g., for two spin-1/2 systems, that $\hat
{S}_{1z} +\hat {S}_{2z} $ is a two-by-two matrix in a chosen basis but $\hat
{S}_{1z}  \hat {S}_{2z} $ is a four-by-four matrix; and
(5) The Hamiltonian of the system must be known in order to construct a matrix for an operator other than the Hamiltonian operator.

\subsection{Difficulties involving the uncertainty principle}

The uncertainty principle is a foundational principle in quantum mechanics and is due to the incompatibility of operators corresponding to observables. In particular, if the operators corresponding to two observables do not commute, there will be an uncertainty relation between them. For example, the uncertainty principle between position and momentum is a particular example of the generalized uncertainty principle and says that the product of the standard deviations in the measurement of position and momentum for a given state of the system (wavefunction) must be greater than or equal to $\hbar/2$. Students have great difficulty with the uncertainty principle. Some major reasons for the difficulty are due to students' misunderstanding of the word uncertainty in this context. In particular, students often incorrectly associate the uncertainty principle with measurement errors or they mistake the concepts of standard deviations and average values (e.g., of position and momentum in the case of the position and momentum uncertainty principle). For example, in one study~\cite{tutorial}, many students incorrectly claimed that when a particle is moving fast, the position measurement has uncertainty because one cannot determine the particle's position precisely. They used this incorrect reasoning to infer that the uncertainty principle is due to the fact that if the particle has a large speed, the position measurement cannot be very precise. 
This type of reasoning is incorrect because it is not the speed of the particle, but rather, the uncertainty in the particle's speed that is related to the uncertainty in position.  
Further discussions with some students with these types of responses indicate that they were confused about measurement errors and/or attributed the uncertainty principle to something related to the expectation values of the different observables. 
In another multi-university study, students were asked a question about position and momentum uncertainty~\cite{zhuqms1}. Many students incorrectly claimed that according to the uncertainty principle, the uncertainty in position is smaller when the expectation value of momentum is larger. Others incorrectly claimed that the expectation value of position is larger when the expectation value of momentum is smaller.
 
\subsection{Difficulties with Dirac notation and issues related to quantum mechanics formalism}

Because Dirac notation is used so extensively in upper-level quantum mechanics, it is important that  students have a thorough understanding of this notation. However, research suggests that students have great difficulties with it~\cite{singh2001,marshman2013}. Below, we give examples of some difficulties found via research.\\

\textbf{Difficulties in consistently recognizing the position space wavefunction in Dirac notation}:
In an investigation on students' facility with this notation, students displayed inconsistent reasoning in their responses to consecutive questions~\cite{marshman2013}. For example, on a multiple-choice survey, three consecutive conceptual questions were posed about the quantum mechanical wave function in position representation, with and without Dirac notation. In the first question, almost all of the students correctly noted that the position space wave function is
$\Psi(x)=\langle x \vert \Psi\rangle$. The second question asked about a generic quantum mechanical operator $\hat Q$ (which is diagonal in the position representation) acting on the state
$\vert \Psi \rangle$  in the position representation, i.e., $\langle x \vert \hat Q \Psi\rangle$. Two of the answer choices were $\hat Q(x)\Psi(x)$  and 
$ \hat Q(x)\langle x \vert \Psi\rangle$, which are both correct since $\Psi(x)=\langle x \vert \Psi\rangle$. However, more than one-third of the students incorrectly claimed that only one of the answers ($ \hat Q(x)\Psi(x)$  or $ \hat Q(x)\langle x \vert \Psi\rangle$) is correct, but not both.  In the third question, more than a third of the students claimed that  ``$\langle x \vert \Psi\rangle=\int_{-\infty}^{+\infty} x \Psi(x) dx$'' is correct. However, it is incorrect because if $\Psi(x)=\langle x \vert \Psi\rangle$, then ``$\langle x \vert \Psi\rangle=\int_{-\infty}^{+\infty} x \Psi(x) dx$'' does not make sense. In a fourth consecutive question, more than a third of the students claimed that $\langle x \vert \Psi\rangle= \int_{-\infty}^{+\infty} \delta(x-x^\prime) \Psi(x^\prime) dx^\prime$ is incorrect. However, it is a correct equality because the integral results in $\Psi(x)=\langle x \vert \Psi\rangle$. We note that the integrals of the type shown above are easy for an advanced student taking quantum mechanics if the problem is given as a math problem without the quantum mechanics context.\\

\textbf{Difficulties with the probability of obtaining a particular outcome for the measurement of an observable in Dirac notation}:
Students also struggled to find the probability of obtaining a particular outcome for a measurement of an observable in a given quantum state when they were asked the question in Dirac notation, even when they correctly identified the same probability in position representation (not written in Dirac notation)~\cite{marshman2013}. For example, in one question, they were told that an operator $\hat Q$ corresponding to a physical observable $Q$ has a continuous non-degenerate spectrum of eigenvalues and the states {$\vert q \rangle$} are the eigenstates of $\hat Q$ with eigenvalues $q$. They were also told that at time $t=0$,  the state of the system is $\vert \Psi\rangle$ and asked to select correct expressions for the probability of obtaining an outcome between $q$ and $q+dq$ if they measure $Q$ at time $t=0$.
The probability of obtaining an outcome between $q$ and $q+dq$ can be written as $\vert \langle q\vert \Psi \rangle \vert^2 dq$ or $\vert \int_{-\infty}^\infty \psi^\star_q(x) \Psi(x)dx \vert^2 dq$ in which $\psi_q(x)$ and $\Psi(x)$ are the wavefunctions corresponding to the states $\vert q \rangle$ and $\vert \Psi \rangle$ respectively.
Some students thought that only the first expression is correct while others claimed that only the second expression is correct.
Pertaining to this issue, one common difficulty revealed in the interviews was related to confusion about projection of a state vector. Projecting state vector $\vert \Psi \rangle$ along an eigenstate $\vert q \rangle$ or a position eigenstate $\vert x \rangle$ gives the probability density amplitude for measuring an eigenvalue $q$ or probability density amplitude for measuring  $x$, respectively, in a state $\vert \Psi \rangle$. These students often incorrectly claimed that an expression for the probability of measuring an observable in an infinitesimal interval must involve integration over $q$ or $x$ even when written in the Dirac notation. \\

\textbf{Difficulties with expectation value, measured values, and their probabilities in Dirac notation}:
In a multi-university study~\cite{singh2001}, students were asked to find a mathematical expression for
$\langle \phi \vert \hat Q \vert \phi \rangle$, where $\vert \phi \rangle$ is a general state and
 the eigenvalue equation for an operator $\hat  Q$ is given by $\hat Q \vert \psi_i \rangle=\lambda_i \vert \psi_i 
\rangle$, $i=1,...,N$.
The correct response is the following: $\langle \phi \vert \hat Q \vert \phi \rangle = \sum_i  \vert \langle \psi_i  \vert \phi \rangle \vert^2 \lambda_i$, or simply
$\sum_i  \vert C_i \vert^2 \lambda_i$, where $C_i=\langle \psi_i  \vert \phi \rangle$. Less than half of the students provided the correct response.
Some students had difficulty with the principle of linear superposition and with Dirac notation. They could not expand a general state in terms of the complete set of eigenstates of an operator. The common mistakes include writing incorrect expressions such as ``$\vert \phi \rangle = \vert \psi \rangle$,'' ``$\vert \phi \rangle = \sum_i \vert \psi_i \rangle$,'' ``$ \langle \phi \vert \psi_i \rangle=1$,'' writing ``$\lambda$'' without any subscript in the answers, making mistakes with summation indices, etc.
Also, many students in the written test and interviews could retrieve from memory that a general
state $\vert \phi  \rangle$ can be expanded as
$\sum_n C_n \vert \psi_n \rangle$ but thought that $\langle \phi \vert \psi_n \rangle$ is unity. This dichotomy suggests that
many students lack a clear understanding of what the expansion $\vert \phi \rangle=\sum_n C_n \vert \psi_n \rangle$ means and
that $C_n=\langle  \psi_n \vert \phi \rangle$ (which implies $\langle \phi \vert \psi_n \rangle=C_n^\star$). In addition, some students thought that the eigenvalue $\lambda_i$ gives the probability of obtaining a particular eigenstate
and expanded the
state as ``{$ \vert \phi \rangle = \sum_i \lambda_i \vert \psi_i \rangle$}.'' 
\\

\textbf{Other difficulties with Dirac notation}:
In the investigation described in Ref.~\cite{marshman2013}, some students also incorrectly claimed that one can always exchange the bra and ket states in the Dirac notation without changing its value if the operator sandwiched between them is a Hermitian operator corresponding to an observable, i.e., $\langle x \vert \hat Q \vert \Psi \rangle=\langle \Psi\vert \hat Q \vert x \rangle$ if $\hat Q$ is Hermitian.  While some of them correctly reasoned that the eigenvalues of a Hermitian operator are real, they erroneously concluded that this implies that one can  exchange the bra and ket states without complex conjugation if the scalar product involves sandwiching a Hermitian operator. 
Students also had difficulties explaining why the scalar product $\langle \Psi \vert \Psi \rangle=1$ is dimensionless whereas $\langle x \vert \Psi \rangle$, which is also a scalar product of two states, has the dimensions of square root of inverse length. Moreover, similar to the difficulties with the position space wavefunction, students also had difficulties with the momentum space wavefunction.

\section{Inadequate problem-solving, reasoning, and self-monitoring skills}

Although the studies discussed so far focused on the difficulties with specific topics while solving non-algorithmic problems, they also reveal that students in upper-level quantum mechanics courses often have inadequate problem-solving, reasoning, and self-monitoring skills. For example, some of these studies show that many students are inconsistent in their reasoning about a particular topic in quantum mechanics across different problems. Their responses are often context dependent and they are unable to transfer their learning from one situation to another appropriately. They often overgeneralize concepts learned in one situation to another in which they are not applicable. They also have difficulty distinguishing between related concepts and often make use of memorized facts and algorithms to solve problems.  Moreover, they often have difficulty solving multi-part problems. The theoretical frameworks discussed earlier suggest that instructors must not only know students' difficulties with various topics, but also their current level of expertise in problem-solving, reasoning, and self-monitoring in order to tailor instruction and build on these skills. For example, according to Hammer's resource model, students' resources include not only their content knowledge, but also the skills they bring to bear to solve problems~\cite{hammer}. To tailor instruction appropriately, instructors should take into account students' resources effectively. In the same manner, in order to provide students an ``optimal mismatch''~\cite{piaget} or to help students remain in the ``zone of proximal development''~\cite{vygotsky}, instruction must build on students' initial knowledge and skills in order to ``stretch'' their learning and develop useful skills. Similarly, to help students remain in the ``optimal adaptibility corrider'' as suggested by Bransford and Schwartz~\cite{schwartz}, students must be given tasks that are appropriate to their skill level and are neither too efficient nor too innovative. All of these theoretical frameworks point to the fact that instructors must not only know students' difficulties with content, but also their level of expertise in their problem-solving, reasoning, and self-monitoring skills in order to help them learn effectively. An understanding of these student difficulties can enable instructors to design instruction to help students learn quantum mechanics content while developing their problem-solving, reasoning, and self-monitoring skills.         

While the studies discussed so far have focused explicitly on investigating students' difficulties with various topics in upper-level quantum mechanics, fewer studies have focused explicitly on students' problem-solving and self-monitoring skills. 
The following two studies~\cite{mason1,shihyin} shed light on the  problem-solving and self-monitoring skills of students in upper-level quantum mechanics.

\subsection{Difficulties with categorizing quantum physics problems}

Categorizing or grouping together problems based upon similarity of solution is often considered a hallmark of expertise. Chi et al.~\cite{chicategorization}  used a categorization task to assess introductory students' expertise in physics. Unlike experts who categorized problems based on the physics principles, introductory students categorized problems involving inclined planes in one category and pulleys in a separate category. Lin et al.~\cite{shihyin}  extended this type of study and performed an investigation in which physics professors and students from two traditionally taught junior/senior level quantum mechanics courses were asked to categorize 20 quantum mechanics problems based upon the similarity of the solution. Professors' categorizations were overall rated higher than those of students by three faculty members who evaluated all of the categorizations without the knowledge of whether those categories were created by the professors or students. The distribution of scores obtained by the students on the categorization task was more or less evenly distributed with some students scoring similar to the professors while others obtained the lowest scores possible. This study suggests that there is a wide distribution in students' performance on a quantum mechanics categorization task, similar to the diversity in students' performance on a categorization of introductory physics problems. Therefore, the study suggests that it is inappropriate to assume that, because they 
 have made it through the introductory and intermediate physics courses, all students in upper-level quantum mechanics will develop sufficient expertise in quantum mechanics after 
traditional instruction. In fact, the diversity in student performance in categorization of quantum mechanics problems suggests that many students are getting distracted by the ``surface features'' of the problem and have difficulty recognizing the deep features which 
 are related
 to how to solve the problem. The fact that many students are struggling to build a robust knowledge structure in a traditionally taught quantum mechanics course suggests that it is inappropriate to assume that teaching by telling is effective for most of these students because it worked for the professors when they were students.

\subsection{Not using problem solving as a learning opportunity automatically}

Reflection and sense-making are integral components of expert behavior. Experts monitor their own learning. They use problem solving as an opportunity for learning, extending, and organizing their knowledge. One related attribute of physics experts is that they learn from their own mistakes in solving problems. Instructors often take for granted that advanced physics students will learn from their own mistakes in problem solving without explicit prompting or incentive, especially if students are given access to clear solutions. It is implicitly assumed  that, unlike introductory students, advanced physics students have become independent learners and will take the time to learn from their mistakes--even if the instructors do not reward them for correcting them, for example, by explicitly asking them to turn in, for course credit, a summary of the mistakes they made and writing how those mistakes can be corrected. Mason et al.~\cite{mason1,mason2} investigated whether advanced students in quantum mechanics have developed these self-monitoring skills and the extent to which they learn from their mistakes. They administered four problems in the same semester twice, both in the midterm and final exams, in a junior/senior level quantum mechanics course. The performance on the final exam shows that while some students performed equally well or improved compared to their performance on the midterm exam on a given question, a comparable number performed poorly both times or regressed (performed well on the midterm exam but performed poorly on the final exam). The wide distribution of students' performance on problems administered a second time points to the fact that many advanced students may not automatically exploit their mistakes as an opportunity for repairing, extending, and organizing their knowledge structure. Mason et al.~\cite{mason1,mason2} also conducted individual interviews with a subset of students to delve deeper into students' attitudes towards learning and the importance of organizing knowledge. They found that some students focused on selectively studying for the exams and did not necessarily look at the solutions provided by the instructor for the midterm exams to learn partly because they did not expect those problems to show up again on the final exam and found it painful to confront their mistakes.

\section{Implications of the research on student difficulties}
The research on student difficulties summarized here can help instructors, researchers, and curriculum designers  design approaches to help students improve their content knowledge and skills and develop a functional understanding of upper-level quantum mechanics. These research studies can also pave the way for future research directions.

\subsection{Research-based Instructional Approaches to Reduce Student Difficulties}

The scaffolding supports that are currently prevalent in research on upper-level quantum mechanics learning involve approaches similar to those that have been found successful at the introductory level~\cite{lillian2,heller,mazur,crouch}. These tools and approaches include: 1) tutorials~\cite{tutorial,sga,measure2,angular2}, which provide a guided inquiry approach to learning; 2) peer-instruction tools~\cite{peer} such as reflective problems and concept-tests, which have been very effective in the introductory physics courses; 3) collaborative problem solving; and 4) kinesthetic explorations~\cite{gire1,gire2}.

Several Quantum Interactive Learning Tutorials (QuILTs) that use a guided inquiry-based approach to learning have been developed to reduce student difficulties~\cite{singhtoday,sga,mz1,mz2,guang2,measure2,ben2014,angular2,tutorial}. They are based on systematic investigations of difficulties students have in learning various concepts in quantum physics. They consistently keep students actively engaged in the learning process by asking them to predict what should happen in a particular situation and then providing appropriate feedback. They often employ visualization tools to help students build physical intuition about quantum processes. QuILTs help students develop content knowledge and skills by attempting to bridge the gap between the abstract, quantitative formalism of quantum mechanics and the qualitative understanding necessary to explain and predict diverse physical phenomena. 
They can be used by instructors in class to supplement lectures. Several students can work on them in groups. QuILTs consist of self-sufficient modular units that can be used in any order that is convenient. The development of a QuILT goes through a cyclical, iterative
process which includes the following stages: (1) development of a preliminary version based on a theoretical analysis of the underlying
knowledge structure and research on student difficulties; (2) implementation and evaluation of the QuILT by administering it individually to students;
(3) determining its impact on student learning and assessing what difficulties were not remedied to the extent desired; and (4) refinements and modifications based on the
feedback from the implementation and evaluation. The topics of these QuILTs include the time-dependent and time-independent Schr\"odinger equation, the time-development of the wave function, the time-dependence of an expectation value, quantum measurement, expectation values, bound and scattering state wave functions, the uncertainty principle, which-path information and double-slit experiments, a Mach-Zehnder interferometer (including the delayed choice experiment, interaction free measurement, quantum eraser, etc.), Stern Gerlach experiments, Larmor precession of spin, quantum key distribution (distribution of a key over a public-channel for encoding and decoding information using single photon states), the basics of a single spin system, and product space and addition of angular momentum (two separate QuILTs on coupled representation and uncoupled representation).
 
A pedagogical approach that has been used extensively in introductory physics courses is peer instruction~\cite{mazur,crouch}. Similar approaches have been effective in helping students learn quantum mechanics~\cite{peer}. In this approach, the instructor poses conceptual, multiple-choice questions to students periodically during the lecture. The focal point of the peer instruction method is the discussion among students based on the conceptual questions. The instructor polls the class after peer interaction to learn about the fraction of students with the correct answer and the types of incorrect answers that are common. Students learn about the course goals and the level of understanding that is desired by the instructor. The feedback obtained by the instructor is also valuable because the instructor determines the fraction of the class that has understood the concepts at the desired level. This peer instruction strategy helps students both learn content knowledge since students must answer conceptual questions and also develop reasoning and self-monitoring skills by asking them to explain their answers to their peers. The method keeps students actively engaged in the learning process and allows them to take advantage of each other's strengths. It helps both the low and high performing students at a given time because explaining and discussing concepts with peers helps even the high performing students organize and solidify concepts in their minds. Recent data suggests that the peer instruction approach is effective in quantum mechanics~\cite{peer}. 

Moreover, for introductory physics, Heller et al.~\cite{heller} have shown that collaborative problem solving is valuable for learning physics and for developing effective problem-solving strategies. Prior research~\cite{coconstruction} has shown that even with minimal guidance from the instructors, introductory physics students can benefit from peer collaboration. In that study, students who worked with peers on conceptual electricity and magnetism questions not only outperformed an equivalent group of students who worked alone on the same task, but collaboration with a peer led to ``co-construction'' of knowledge in $29\%$ of the cases.  Co-construction of knowledge occurs when neither student who engaged in peer collaboration was able to answer the questions before the collaboration, but both were able to answer them after working with a peer on a post-test given individually to each person. Similar to the introductory physics study involving co-construction~\cite{coconstruction}, a study was conducted in which conceptual questions on the formalism and postulates of quantum mechanics were administered individually and in groups of two to 39 upper-level students. It was found that co-construction occurred in $25\%$ of the cases in which both students individually had selected an incorrect answer~\cite{marshman2015}. 
  
Developing a functional knowledge is closely connected to having appropriate epistemological views of the subject matter. Epistemological beliefs can affect students' motivation, enthusiasm to learn, time on task, approaches to learning, and ultimately, learning. Motivation can play a critical role in students' level and type of cognitive engagement in learning quantum mechanics. What types of instructional strategies can help improve students' epistemological views? Similar to students' views about learning in introductory mechanics, students' epistemological views about learning quantum mechanics can be improved if instructional design focuses on sense making and learning rather than on memorization of facts and accepting the instructor as authority. These effective instructional strategies should include encouraging students to work with peers to make sense of the material and providing problems in contexts that are interesting and appealing to students. Kinesthetic explorations~\cite{gire1,gire2} can also be effective in this regard. Both formative assessments (e.g., peer instruction with concept tests, tutorial pre-tests/post-tests,  collaborative problem-solving, homework assignments) and summative assessments (e.g., exams) should include  problems that help students with conceptual reasoning and sense-making. Problems involving interesting applications such as quantum key distribution, Mach-Zehnder interferometer with single photons, and quantum eraser can be beneficial. Otherwise, students may continue to perform well on exams without developing a functional understanding, e.g., by successfully solving algorithmic problems involving solutions of the time-independent Schr\"odinger equation with complicated boundary conditions and potential energies.

\subsection{Concluding Remarks and Future Directions}
 
Mathematically skilled students in a traditional introductory physics course focusing on mastery of algorithms can ``hide'' their lack of conceptual knowledge behind their mathematical skills~\cite{mazur}. However, their good performance on algorithmic physics problems does not imply that they have engaged in self-regulatory activities throughout the course or have built a hierarchical knowledge structure. In fact, most physics faculty, who teach both introductory and advanced courses, agree that the gap between conceptual and quantitative learning gets wider in a traditional physics course from the introductory to advanced level. Therefore, students in a traditionally taught and assessed quantum mechanics course can hide their lack of conceptual knowledge behind their mathematical skills even better than students in introductory physics. Closing the gap between conceptual and quantitative problem-solving by assessing both types of learning is essential to helping students in quantum mechanics develop functional knowledge. Interviews with faculty members teaching upper-level quantum mechanics~\cite{dubson,siddiqui} suggest that some assign only quantitative problems in homework and exams (e.g., by asking students to solve the time-independent Schr\"odinger equation with complicated boundary conditions) because they think students will learn the concepts on their own. Nevertheless, as illustrated by the examples of difficulties in this paper, students may not  adequately learn about quantum mechanics concepts unless course assessments value conceptual learning, sense-making, and the building of a robust knowledge structure. Therefore, to help students develop a functional knowledge of quantum mechanics, formative and summative assessments should emphasize the connection between conceptual understanding and mathematical formalism. 

Further research comparing traditional and transformed upper-level quantum mechanics courses should be conducted to shed light on the extent to which students are making an effort to extend, organize, and repair their knowledge structure and develop a functional understanding. It would be valuable for future research studies to also investigate the extent to which students in these courses are making a connection between mathematics and physics, whether it is to interpret the physical significance of mathematical procedures and results, convert a real physical situation into a mathematical model, or apply mathematical procedures appropriately to solve the physics problems beyond memorization of disconnected pieces for exams. Students' ability to estimate physical quantities and evaluate limiting cases in different situations as appropriate and their physical intuition for the numbers across different content areas in traditional and transformed courses can be useful for evaluating their problem-solving, reasoning, and metacognitive skills. Although tracking the same student's learning and self-monitoring skills longitudinally is a difficult task, taking snapshots of physics majors' learning and self-monitoring skills across different physics content areas and across contexts within a topic can be very valuable. It would also be useful to explore the impact of traditional and non-traditional homework (e.g., reflective problems which are conceptual in nature) on student learning.   More research on traditional and transformed courses is also needed to investigate the facility with which upper-level students transfer what they learned in one context to another context in the same course, whether students retain what they have learned when the course is over, and whether they are able to transfer their learning from one course to another (e.g., from quantum mechanics to statistical mechanics) or whether such transfer is rare. It will be useful to investigate the types of scaffolding supports that may significantly improve students' problem-solving, reasoning, and metacognitive skills in upper-level quantum mechanics and how and when such support should be decreased. 
   
Finally, research should also focus on how community building affects how students learn quantum mechanics and on effective strategies for making students part of a learning community. It will also be useful to investigate the quality of students' communication about course content with their peers and the instructor in transformed upper-level quantum mechanics courses and learn about the extent to which students are more advanced compared to introductory physics students in the level of sophistication displayed by their word usage, terminology, and related semantics. We hope that this review of student difficulties will be helpful for developing learning tools and approaches to improve student learning of quantum mechanics.

\begin{theacknowledgments}

This work is supported in part by the National Science Foundation awards PHY-0968891 and PHY-1202909.
We are very grateful to all faculty members and students who helped with these investigations. We thank F. Reif, R. P. Devaty and all of the members of the Physics Education Research group at the University of Pittsburgh for helpful discussions and constructive feedback.

\end{theacknowledgments}

\bibliographystyle{aipproc}   

\begin{thebibliography}{9}
\bibitem{arons} For example, see A. B. Arons, {\it A Guide to Introductory Physics Teaching} (John Wiley \& Sons, NY, 1990).
\bibitem{lillian} L. C. McDermott, Research on conceptual understanding in mechanics, Phys. Today \textbf{37} (7), 24-32 (1984).
\bibitem{mcdermott1} L. McDermott, P. Shaffer, and M. Somers, Research as a guide for teaching introductory mechanics: An illustration in the 	context of the Atwood's machine, Am. J. Phys. \textbf{62}, 46 (1994).
\bibitem{mcdermott2} L. McDermott, Oersted Medal Lecture 2001: Physics education research-The key to student learning, Am. J. Phys. \textbf{69}, 1127 	(2001).
\bibitem{reif} F. Reif, J. Larkin, and G. Brackett, Teaching general learning and problem-solving skills, Am. J. Phys. \textbf{44}, 212 (1976).
\bibitem{reif2} F. Reif, Instructional design, cognition, and technology: Applications to the teaching of scientific concepts, Journal of Research in Science Teaching \textbf{24}, 309 (1987).
\bibitem{singh2001} C.  Singh, Student understanding of quantum mechanics, Am. J. Phys. \textbf{69} (8), 885--896 (2001). 
\bibitem{loverude} M. Loverude, C. Kautz, and P. Heron, Student understanding of the first law of thermodynamics: Relating work to the adiabatic 	compression of an ideal gas, Am. J. Phys. \textbf{70}, 137 (2002).
\bibitem{chasteen} S. Chasteen, S. Pollock, R. Pepper, and K. Perkins, Transforming the junior level: Outcomes from instruction and research in electricity and magnetism, Phys. Rev. ST PER \textbf{8}, 020107 (2012).
\bibitem{simon} A. Newell and H. Simon, {\it Human Problem-solving} (Prentice Hall, NJ, 1972).
\bibitem{anderson} J. R. Anderson, {\it Learning and Memory: An Integrative Approach} (John Wiley \& Sons, NY, 1999), 2nd ed.
\bibitem{ericsson} K. Ericsson and J. Smith, {\it Toward a General Theory of Expertise: Prospects and Limits} (Cambridge, UK, 1991).
\bibitem{mason1} A. Mason and C. Singh, Do advanced students learn from their mistakes without explicit intervention?, Am. J. Phys. \textbf{78} (7), 760 (2010).
\bibitem{shihyin} S. Y. Lin and C. Singh, Categorization of quantum mechanics problems by professors and students, Euro. J. Phys. \textbf{31}, 57 (2010); S. Y. Lin and C. Singh, Assessing Expertise in Quantum Mechanics using Categorization Task,  Proceedings of the Phys. Ed. Res. Conference, Ann Arbor, MI, (M. Sabella, C. Henderson, C. Singh Eds.), AIP Conf. Proc., Melville, New York \textbf{1179}, 185-188 (2009).
\bibitem{metacognition} G. Schraw, Promoting general metacognitive awareness, Instructional Science \textbf{26}, 113 (1998).
\bibitem{kuhn} T. Kuhn, {\it The Structure of Scientific Revolutions} (University of Chicago, Chicago, 1962).
\bibitem{griffiths} D. J. Griffiths, \textit{Introduction to Quantum Mechanics} (Prentice Hall, Upper Saddle River, NJ, 1995). 
\bibitem{disessa1} A.  diSessa, Knowledge in pieces, in {\it Constructivism in the Computer Age}, edited by G. Forman and P. Pufall (Lawrence 	Erlbaum, Hillsdale, NH, 1988), Chap. 4, pp. 49-70.
\bibitem{disessa2} A. diSessa, What changes in conceptual change? International Journal of Science Education \textbf{20}, 1155 (1998).
\bibitem{brad} B. Ambrose, Ph.D. Thesis, University of Washington (1999).
\bibitem{sadaghiani} H. Sadaghiani, Ph.D. Thesis, The Ohio State University (2005).
\bibitem{guangtian} Guangtian Zhu, Ph.D. Thesis, University of Pittsburgh (2011).
\bibitem{singhtoday} C. Singh, M. Belloni, and W. Christian, Improving student's understanding of quantum mechanics,
Physics Today {\bf 8}, 43 (2006).
\bibitem{singhgrad} C. Singh, Student understanding of quantum mechanics at the beginning of graduate instruction, Am. J. Phys. \textbf{76} (3), 277 (2008).
\bibitem{aip} AIP Statistical Research Center, www.aip.org/statistics
\bibitem{gick} M. L. Gick and K. J. Holyoak, The cognitive basis of knowledge transfer, in {\it Transfer of Learning}, edited by S.M. Cormier and J.D. Hagman (Academic Press, NY, 1987), pp. 212-231.

\bibitem{zollman} D. Zollman, S. Rebello, and K. Hogg, Quantum physics
for everyone: Hands-on activities integrated with technology, Am. J. Phys.
\textbf{70} (3), 252 (2002).
\bibitem{zollman2} P. Jolly, D. Zollman, S. Rebello, and A. Dimitrova, Visualizing potential energy diagrams, Am. J. Phys. {\bf 66} (1), 57 (1998). 
\bibitem{redish} L. Bao and E. Redish, Understanding probabilistic
interpretations of physical systems: A prerequisite to learning quantum
physics, Am. J. Phys. \textbf{70} (3), 210 (2002).
\bibitem{redish2}  M. Wittmann, R. Steinberg, and E. Redish, Investigating student understanding of quantum physics: Spontaneous models of conductivity, Am. J. Phys. \textbf{70} (3), 218 (2002).
\bibitem{narst} \textit{Research on Teaching and Learning of
Quantum mechanics}, papers presented at the National Association
for Research in Science Teaching, \url{<perg.phys.ksu.edu/papers/narst/>} (1999).
\bibitem{bailey} C. Baily and N. Finkelstein, Teaching and understanding of quantum interpretations in modern physics courses, Phys. Rev. ST PER {\bf{6}}, 010101 (2010). 
\bibitem{fischler} H. Fischler and M. Lichtfeldt, Modern physics and
students' conceptions, Int. J. Sci. Educ. \textbf{14} (2), 181--190 (1992).
\bibitem{ireson} G. Ireson, The quantum understanding of pre-university physics students, Phys. Educ. {\bf 35} (1), 15--21 (2000).
\bibitem{niedderer} J. Petri and H. Niedderer, A learning pathway in high-school level
quantum atomic physics, Int. J. Sci. Educ. {\bf 20} (9), 1075--1088 (1998). 
\bibitem{muller} R. Muller and H. Wiesner, Teaching quantum mechanics on an introductory level, Am. J. Phys. \textbf{70}, 200--209 (2002).
\bibitem{lawless} C. Lawless, Investigating the cognitive structure of students studying quantum theory in an open university history of science course: A pilot study, British J. Ed. Tech. {\bf 25} (3), 198--216 (1994).
\bibitem{mckagan} S. McKagan, K. Perkins and C. Wieman, 
Reforming a large lecture modern physics course for engineering majors using a PER-based design, in {\it Proceedings of the 2006 Phys. Ed. Res. Conf., Syracuse, NY}, edited by L. McCullough, L. Hsu, and P. Heron (AIP, Melville, NY, 2006) pp. 34-37. 
\bibitem{photo} S. McKagan, W. Handley, K. Perkins and C. Wieman, 
A research-based curriculum for teaching the photoelectric effect, Am. J. Phys. {\bf 7}, 87 (2009).
\bibitem{bohrmodel} S. McKagan, K. Perkins and C. Wieman, 
Why we should teach the Bohr model and how to teach it effectively,
Phys. Rev. ST PER {\bf 4}, 010103 (2008).
\bibitem{phet} S. McKagan, K. Perkins, M. Dubson, C. Malley, S. Reid, R. LeMaster, and C. Wieman, Developing and researching PhET simulations for teaching quantum mechanics, Am. J. Phys. {\bf 76}, 406 (2008).
\bibitem{tunnel} S. McKagan, K. Perkins, and C. Wieman, 
A deeper look at student learning of quantum mechanics: The case of tunneling,
Phys. Rev. ST PER {\bf 4}, 020103 (2008).
\bibitem{qmcs} S. B. McKagan and C. E. Wieman, 
Exploring student understanding of energy through the quantum mechanics conceptual survey, in {\it Proceedings of the 2005 Physics Education Research Conference, Salt Lake City, UT}, edited by P. Heron, L. McCullough, and J. Marx (AIP, Melville, NY, 2006) pp. 65-68.
\bibitem{visual} For example see http://web.phys.ksu.edu/vquantum mechanics/ or simulations available at \url{<www.physics.umd.edu/perg/quantum mechanics/quantum mechanicscourse/NewModel>}.
\bibitem{robinett} E. Cataloglu and R. W. Robinett, 
Testing the development of student conceptual and visualization skills in 
quantum mechanics through the undergraduate career, Am. J. Phys. {\bf 70}, 238--251 (2002).
\bibitem{mario} See for example, see \url{<www.opensourcephsyics.org>}.
\bibitem{physlet}
M. Belloni and W. Christian, Physlets for quantum mechanics,
Comp. Sci. Eng. {\bf 5} (1), 90-96 (2003);
M. Belloni, W. Christian and A. Cox, {\it Physlet Quantum Physics} (Pearson Prentice Hall, Upper Saddle River, NJ, 2006).
\bibitem{brandt}S. Brandt and H. Dahmen, \textit{The Picture Book of Quantum
Mechanics} (Springer-Verlag, New York, 2001).
\bibitem{thaller}B. Thaller, \textit{Visual Quantum Mechanics} (Springer-Verlag, New York, 2000).
\bibitem{hiller}J. Hiller, I. Johnston, and D. Styer,
\textit{Quantum Mechanics Simulations} (John Wiley \& Sons, New York, 1995).
\bibitem{antje} A. Kohnle et al., A new introductory quantum mechanics curriculum, Eur. J. Phys. 35, 015001 (2014); http://www.st-andrews.ac.uk/physics/quvis
\bibitem{johnson} I. D. Johnston, K. Crawford, and P. R. Fletcher, Student difficulties in learning quantum mechanics, Int. J. Sci. Educ. \textbf{20}, 427 (1998).
\bibitem{passante} G. Passante, P. Emigh and P. Shaffer, Investigating student understanding of basic quantum mechanics in the context of time-dependent perturbation theory, in {\it Proceedings of the 2013 Phys. Ed. Res. Conference, Portland, OR}, edited by A. Churukian, P. Engelhardt, D. Jones (2014) pp. 269-272 (doi 10.1119.perc.2013.pr.010).

\bibitem{gire1} E. Gire and C. Manogue, Making sense of operators, eigenstates, and quantum measurements, in {\it Proceedings of the 2011 Physics Education Research Conference, Omaha, NE}, edited by N. Rebello, P. Engelhardt, and C. Singh (AIP, Melville, NY, 2011) pp. 195-198.
\bibitem{hammer} D. Hammer, Epistemological beliefs in introductory physics, Cognition and Instruction \textbf{12}, 151 (1994).
\bibitem{piaget} J. Piaget, {\it Success and Understanding} (Harvard University Press, Cambridge, MA, 1978).
\bibitem{posner} G. Posner, K. Strike, W. Hewson, and W. Gertzog, Accomodation of a scientific conception: Toward a theory of conceptual 	change, Science Education. \textbf{66}, 211 (1982).
\bibitem{schwartz} D. Schwartz, J. Bransford, and D. Sears, Efficiency and innovation in transfer, in {\it Transfer of learning from a modern 	multidisciplinary perspective}, edited by J. Mestre (Information Age, 2005) pp.1-51.  
\bibitem{vygotsky} L. Vygotsky, {\it Mind in Society: The Development of Higher Psychological Processes} (Harvard University Press, 1978).
\bibitem{chi} M.T.H. Chi, in \textit{ The Thinking Aloud Method}, edited by M. W. van Someren, Y. F. Barnard and J.A.C. Sandberg 
(Academic Press, London, 1994) Chap. 1.
\bibitem{analogous} C. Singh and E. Marshman, Analogous patterns of student reasoning difficulties in introductory physics and upper-level quantum mechanics,  in {\it Proceedings of the 2013 Phys. Ed. Res. Conference, Portland, OR}, edited by A. Churukian, P. Engelhardt, and D. Jones (2014) pp. 46-49 (doi 10.1119.perc.2013.pr.010).
\bibitem{wittman1} M. Wittmann, J. Morgan, and L. Bao, Addressing student models of energy loss in quantum tunneling, European Journal of Physics {\bf 26}, 939 (2005).

\bibitem{wittman2} J. Morgan, and M. Wittmann, Examining the evolution of student ideas about quantum tunneling, in {\it Proceedings of the 2005 Phys. Ed. Res. Conference, Salt Lake City, UT}, edited by P. Heron, L. McCullough, and J. Marx (AIP, Melville, NY, 2006) pp. 73-76.
\bibitem{zhuqms1} G. Zhu and C. Singh, Surveying students' understanding of quantum mechanics in one spatial dimension, Am. J. Phys. \textbf{80} (3), 252-259 (2012).
\bibitem{zhuqms} G. Zhu and C. Singh, Surveying students' understanding of quantum mechanics, in {\it Proceedings of the 2009 Phys. Ed. Res. Conference, Portland, OR}, edited by C. Singh, M. Sabella, and S. Rebello (AIP, Melville, NY, 2010) pp. 301-304.
\bibitem{singhtransfer} C. Singh, Transfer of learning in quantum mechanics, in {\it Proceedings of the 2003 Phys. Educ. Res. Conference, Sacramento, CA}, edited by P. Heron, S. Franklin, and J. Marx (AIP, Melville, NY, 2004) pp. 23-26. 
\bibitem{sga} G. Zhu and C. Singh, Improving students' understanding of quantum mechanics via Stern-Gerlach experiment, Am. J. Phys \textbf{79} (5), 499-507, (2011); G. Zhu and C. Singh, Students' Understanding of Stern Gerlach Experiment, Proceedings of the Phys. Ed. Res. Conference, Ann Arbor, MI, (M. Sabella, C. Henderson, C. Singh Eds.), AIP Conf. Proc., Melville, New York textbf{1179}, 309-312 (2009).
\bibitem{my2} C. Singh, Student understanding of quantum mechanics formalism, in {\it Proceedings of the 2006 Phys. Educ. Res. Conference, Syracuse, NY}, edited by L. McCullough, L. Hsu and P. Heron (AIP, Melville, NY, 2007) pp. 185-188.
\bibitem{sethqkd} S. DeVore and C. Singh, Development of an interactive tutorial on quantum key distribution, in {\it Proceedings of the 2014 Phys. Ed. Res. Conference, Minneapolis, MN}, edited by A. Churukian, P. Engelhardt, and D. Jones (2015) pp. 59-62.
\bibitem{mz1} C. Singh and E. Marshman, Developing an interactive tutorial on a Mach-Zehnder interferometer with single photons, in {\it Proceedings of the 2014 Phys. Ed. Res. Conference, Minneapolis, MN}, edited by P. Engelhardt, A. Churukian, and D. Jones (2015) pp. 239-242.

\bibitem{mz2} E. Marshman and C. Singh, Developing an interactive tutorial on a quantum eraser,  in {\it Proceedings of the 2014 Phys. Ed. Res. Conference, Minneapolis, MN}, edited by P. Engelhardt, A. Churukian, and D. Jones (2015) pp. 175-178.
\bibitem{maries} A. Maries and C. Singh, unpublished data.

\bibitem{marshman2013} C. Singh and E. Marshman, Investigating student difficulties with Dirac notation, in {\it Proceedings of the 2013 Phys. Ed. Res. Conference, Portland, OR}, edited by A. Churukian, P. Engelhardt, and D. Jones (2014) pp. 345-348 (doi 10.1119.perc.2013.pr.074).

\bibitem{guang2} C. Singh and G. Zhu, Cognitive issues in learning advanced Physics: An example from quantum mechanics, in {\it Proceedings of the 2008 Phys. Ed. Res. Conference, Ann Arbor, MI}, edited by M. Sabella, C. Henderson, C. Singh (AIP, Melville, NY, 2009) pp. 63-66.
\bibitem{improve} C. Singh, Assessing and improving student understanding of quantum mechanics, in {\it Proceedings of the 2005 Phys. Ed. Res. Conference, Salt Lake City, UT}, edited by P. Heron, L. McCullough and J. Marx (AIP, Melville, NY, 2006) pp. 69-72.
\bibitem{measure01} G. Zhu and C. Singh, Improving students' understanding of quantum measurement, in {\it
Proceedings of the 2009 Phys. Ed. Res. Conference, Portland, OR}, edited by C. Singh, M. Sabella, and S. Rebello (AIP, Melville, NY, 2010) pp. 345-348.
\bibitem{measure02} G. Zhu and C. Singh, Students' difficulties with quantum measurement, in {\it Proceedings of the 2011 Phys. Ed. Res. Conference, Omaha, NE}, edited by S. Rebello, C. Singh, and P. Engelhardt (AIP, Melville, NY, 2012) pp. 387-390.
\bibitem{measure1} G. Zhu and C. Singh, Improving students' understanding of quantum measurement I: Investigation of difficulties, Phys. Rev. ST PER, \textbf{8} (1), 010117 (2012).
\bibitem{measure2} G. Zhu and C. Singh, Improving students' understanding of quantum measurement II: Development of research-based learning tools, Phys. Rev. ST PER, \textbf{8} (1), 010118 (2012).

\bibitem{my1} C. Singh, Helping students learn quantum mechanics for quantum computing, in {\it Proceedings of the 2006 Phys. Educ. Res. Conference, Syracuse, NY}, edited by L. McCullough, L. Hsu and P. Heron (AIP, Melville, NY, 2007) pp. 42-45.
\bibitem{marshmantime} E. Marshman and C. Singh, Investigating student difficulties with time dependence of expectation values in quantum mechanics,  in {\it Proceedings of the 2013 Phys. Ed. Res. Conference, Portland, OR}, edited by P. Engelhardt, A. Churukian and D. Jones (2014) pp. 46-49 (doi 10.1119.perc.2013.pr.049).

\bibitem{ben2014} B. Brown and C. Singh, Development and evaluation of a quantum interactive learning tutorial on Larmor precession of spin,  in {\it Proceedings of the 2014 Phys. Ed. Res. Conference, Minneapolis, MN}, edited by A. Churukian, P. Engelhardt, and D. Jones (2015) pp. 47-50.
\bibitem{angular1} C. Singh and G. Zhu, Students' understanding of the addition of angular momentum, in {\it Proceedings of the 2011 Phys. Ed. Res. Conference, Omaha, NE}, edited by S. Rebello, C. Singh, and P. Engelhardt (AIP, Melville, NY, 2012) pp. 355-358.

\bibitem{angular2} G. Zhu and C. Singh, Improving students' understanding of the addition of angular momentum in quantum mechanics, Phys. Rev. ST PER \textbf{9} (1), 010101 (2013).

\bibitem{tutorial} C. Singh, Interactive learning tutorials on quantum mechanics, Am. J. Phys. \textbf{76}(4), 400 (2008).

\bibitem{chicategorization} M. T. H. Chi, P. J. Feltovich, and R. Glaser, Categorization and representation of physics knowledge by experts and novices,	Cogn. Sci. \textbf{5}, 121 (1981).

\bibitem{mason2} A. Mason and C. Singh, Reflection and self-monitoring in quantum mechanics, in {\it  Proceedings of the 2009 Phys. Ed. Res. Conference, Ann Arbor, MI}, edited by M. Sabella, C. Henderson, and C. Singh, {\bf 1179} (AIP, Melville, NY, 2009) pp. 197-200.

\bibitem{lillian2} L. McDermott and the Physics Education Group at the University of Washington, {\it Physics by Inquiry, Vols. I and II.} (John Wiley $\&$ Sons Inc., New York, NY, 1996).

\bibitem{heller} P. Heller, R. Keith, and S. Anderson, Teaching problem solving through cooperative grouping. Part 1: Group versus individual 	problem solving, Am. J. Phys \textbf{60}, 627 (1992).

\bibitem{mazur} E. Mazur, 
{\it Peer Instruction: A User's Manual} (Prentice Hall, Upper Saddle River, NJ, 1997).

\bibitem{crouch} C. Crouch and E. Mazur, Peer Instruction: Ten years of experience and results, Am. J. Phys. {\bf 69} (9), 970-977 (2001).

\bibitem{peer} C. Singh and G. Zhu, Improving students' understanding of quantum mechanics by using peer instruction, in {\it Proceedings of the 2011 Phys. Ed. Res. Conference, Omaha, NE}, edited by S. Rebello, C. Singh, and P. Engelhardt (AIP, Melville, NY, 2012) pp. 77-80, G. Zhu and C. Singh, Peer Instruction for Quantum Mechanics,  American Physical Society (APS) Forum on Education Newsletter, 8-10, Fall (2009).

\bibitem{gire2} C. Manogue, and E. Gire, Representations for a spins first approach to quantum mechanics, in {\it Proceedings of the 2011 Phys. Ed. Res. Conference, Omaha, NE}, edited by N. Rebello, P. Engelhardt and C. Singh (AIP, Melville, NY, 2012) pp.55-58.

\bibitem{coconstruction} C. Singh, Impact of peer interaction on conceptual test performance, Am. J. Phys. {\bf 73}, 446 (2005).

\bibitem{marshman2015} E. Marshman and C. Singh, unpublished data.

\bibitem{dubson} M. Dubson, S. Goldhaber, S. Pollock, and K. Perkins, 
Faculty disagreement about the teaching of quantum mechanics, in {\it Proceedings of the 2009 Phys. Ed. Res. Conference, Portland, OR}, edited by C. Singh, M. Sabella and S. Rebello (AIP, Melville, NY, 2009) pp. 137-140.

\bibitem{siddiqui} S. Siddiqui and C. Singh, Surveying instructors' attitudes and approaches to teaching quantum mechanics, in {\it  
Proceedings of the 2009 Phys. Ed. Res. Conference, Portland, OR}, edited by C. Singh, M. Sabella, and S. Rebello (AIP, Melville, NY, 2010) pp. 297-300.

\end{thebibliography}

\pagebreak

\end{document}